\documentclass[a4paper, accepted=2020-10-12]{quantumarticle}

\pdfoutput=1

\usepackage[utf8]{inputenc}
\usepackage{tikz} 
\usepackage[pretty,strict,uselistings]{revquantum}
    \newrm{N}

    \newaffil{TRIUMF}{
        TRIUMF, 
        Vancouver, British Columbia, Canada V6T2A3 
    }

    \newaffil{Sandia}{
        Quantum Performance Laboratory, Sandia National Laboratories,
        Albuquerque, New Mexico 87185, USA
    }

    \newaffil{UW}{
        Department of Physics and Astronomy, 
        University of Waterloo, 
        Waterloo, ON, Canada
    }
    
    \newaffil{UWb}{
        Department of Physics, 
        University of Washington, 
        Seattle, WA 98195, USA
    }
    \newaffil{PNNL}{
        Pacific Northwest National Laboratory, 
        Richland, WA 99352, USA
    }
    
\usepackage[sort&compress,numbers,merge]{natbib}
\usepackage[bold]{hhtensor}

\usepackage{bm}
\usepackage[normalem]{ulem}

\newoperator{GL}
\newoperator{Hom}
\newoperator{Bin}
\newrm{ideal}
\newrm{validate}
\newrm{click}
\newrm{op}
\newrm{SPAM}
\newrm{BME}
\newrm{ave}
\newrm{Bayes}

\newtheorem{definition}{Definition}
 
\usepackage{amsmath}
        
\usepackage{nicefrac}
    
\usepackage{appendix}

\newcommand{\sket}[1]{|#1\rangle\!\!\rangle}
\newcommand{\sbra}[1]{\langle\!\!\langle#1|}
\newcommand{\sbraket}[1]{\langle\!\!\langle#1\rangle\!\!\rangle}

\newcommand{\Hil}{\mathcal{H}}
\newcommand{\Lin}{\mathrm{L}}
\newcommand{\Tra}{\mathrm{T}}
\newcommand{\Gua}{\mathrm{G}}
\newcommand{\CC}{\mathbb{C}}

\newcommand{\Seq}{\mathcal{S}}
\newcommand{\BS}{\mathcal{B}}
\newcommand{\GS}{\mathcal{G}}

\newcommand{\figurefolder}{../fig}

\renewcommand{\figurefolder}{fig}
\begin{document}

\title{Operational, gauge-free quantum tomography}
\author{Olivia Di Matteo}
    \affilTRIUMF
    \affilUWPhysAstro
    \affilIQC
\author{John Gamble}
    \affilMSRQuArC
\author{Chris Granade}
    \affilMSRQuArC
\author{Kenneth Rudinger}
    \affilSandia
\author{Nathan Wiebe}
    \affilMSRQuArC
	\affilUWb
	\affilPNNL

\date{authors in alphabetical order by last name}

\begin{abstract}
As increasingly impressive quantum information processors are realized in laboratories around the world, robust and reliable characterization of these devices is now more urgent than ever.
These diagnostics can take many forms, but one of the most popular categories is \emph{tomography}, where an underlying parameterized model is proposed for a device and inferred by experiments.
Here, we introduce and implement efficient operational tomography, which uses experimental observables as these model parameters. 
This addresses a  problem of ambiguity in representation that arises in current tomographic approaches (the \emph{gauge problem}).
Solving the gauge problem enables us to efficiently implement operational tomography in a Bayesian framework computationally, and hence gives us a natural way to include prior information and discuss uncertainty in fit parameters. 
We demonstrate this new tomography in a variety of different experimentally-relevant scenarios, including standard process tomography, Ramsey interferometry, randomized benchmarking, and gate set tomography. 
\end{abstract}

\maketitle
\tableofcontents

\section{Introduction}

Quantum computing offers the potential for significant advantages across a wide range of important problems.
Establishing a rigorous understanding of the costs involved in producing enterprise-scale quantum computers is a critical part of current decision making.
This need has driven efforts to more precisely estimate the costs of different algorithms across different applications, such as in quantum chemistry simulations \cite{rws+_elucidating_2017}.
However, these resource estimations depend critically on the \emph{quality} of the qubits used, \emph{i.e.}, the accuracy with which one can perform quantum gates and measurements.
The collection of procedures used for detecting and debugging faulty operations on quantum computers is known as quantum characterization, verification, and validation (QCVV).
Through QCVV, scientists and engineers working on quantum hardware can hope to diagnose errors and certify performance, in turn improving qubit design and operation.

One goal of QCVV is to learn what actually happens when we attempt to apply a target unitary operator $U$, a procedure broadly known as \emph{quantum tomography}.
Using the language of open quantum systems, we can hypothesize that there is some \emph{channel} $\Lambda$ that, if we knew it, would allow us to predict what happens when we apply $U$ to any state.
The problem then becomes determining how should we best learn $\Lambda$ given experimental evidence from our quantum device.
The tomography problem has been approached in a wide variety of ways ~\cite{fer_selfguided_2014,abj+_ancillaassisted_2003,gcc_practical_2016,
blu_optimal_2010,hh_adaptive_2012,mckay2020correlated,quadeer2019minimax,
cerfontaine2019self,guff2019decision,fiderer2020neural,lukens2020practical}.
The various procedures generally learn $\Lambda$ by (repeatedly) preparing a variety of input states $\{\rho_i\}$, sending each through the application of $U$, and then measuring a variety of effects $\{E_j\}$.
In some cases, the use of auxiliary qubits as a reference can eliminate the need to vary over input states \cite{alt_aapt_2003}. 
However, these latter approaches are mainly useful for reasoning mathematically about 
tomography \cite{gcc_practical_2016} and offer limited experimental applicability.
Hence, we will focus on the more typical case in this work.

While valid, this rests critically on the assumption that we know what each of $\{\rho_i\}$ and $\{E_j\}$ are.
In practice, each state $\rho_i$ and each measurement $E_j$ may be subject to its own physical errors, and these may in turn be objects that we would like to learn.
Worse still, we often prepare states by performing a particular privileged state preparation procedure $\rho_{\star}$, and then applying unitary evolution operators $\{V_i\}$ to obtain $\rho_i \defeq V_i \rho_{\star} V_i^{\dagger}$.
Similarly, measurements are often effected by transforming a particular privileged measurement under unitary evolution.

Once we include the experimental consideration that the channels we would like to study are the same ones that we use to prepare and measure our devices, we are forced to ensure that we learn the channels describing our system in a self-consistent manner.
We cannot learn a channel $\Lambda$ entirely independently of the experimental context in which $\Lambda$ occurs, but must describe that channel such that we can predict its action in a larger experiment.
This self-consistency requirement then forces us to face another difficulty: we can always transform the entire description of an experiment in a consistent fashion, such that there is \emph{no} observable difference whatsoever.
For instance, the states $\ket{0}$ and $\ket{1}$ are in essence just labels for two levels of a quantum system; there is no observable impact to our calling them $\ket{\heartsuit}$ and $\ket{\diamondsuit}$, $\ket{\sharp}$ and $\ket{\flat}$, or even $\ket{1}$ and $\ket{0}$.

That we can rename $\ket{0}$ to $\ket{1}$ and vice versa illustrates one way to formally describe the challenge imposed by self-consistency.
In particular, if we perform an experiment whose outcomes are described by Born's rule as $\Tr(E \Lambda[\rho])$, then for any unitary map $U$ the experiment $\Tr(U E U^\dagger \cdot U \Lambda[U^\dagger \cdot U\rho U^\dagger \cdot U] U^\dagger)$ has the exact same outcome distribution, and thus cannot be distinguished from our original description.
That is, we cannot decide if we have $(\rho, E, \Lambda)$ or if we have $(U\rho U^\dagger, U E U^\dagger, U \Lambda U^\dagger)$.

Taking a step back, something seemingly ridiculous has happened: we asked merely for a description of how one component of our quantum device operates, and arrived at a seemingly fundamental limit to what knowledge we can ever gather about our device.
After all, $U \rho U^\dagger$ and $\rho$ seem to be very different preparation procedures!
Recently, gate set tomography (GST) \cite{mgs+_selfconsistent_2013,bgn+_robust_2013}~has been used as a means to solve this conundrum by explicitly including the effects of this apparent ambiguity into estimation procedures.
With GST, we perform inference on the entire gate set, state preparation, and measurement procedure based on empirical frequencies from repeated experiments. 
This inference procedure can be quite sophisticated in practice, with carefully designed experiments to tease out very slight channel imperfections.  
Over the past several years, GST has been demonstrated experimentally on a wide variety of platforms \cite{bgn+_robust_2013, dehollain2016optimization, BlumeKohoutDemonstrationqubitoperations2017, rudinger2017experimental, rudinger2019probing, rol2017restless, chen2019detector, geller2020rigorous, govia2020bootstrapping, hong2020demonstration, hughes2020benchmarking, joshi2020quantum, mavadia2018experimental, ware2017experimental}, predominately using the software package pyGSTi \cite{nielsen2019python, nielsen2020probing}.

Of course, gate set tomography also has drawbacks, suffering from a conceptual difficulty known as the 
\emph{gauge problem} \cite{BlumeKohoutDemonstrationqubitoperations2017,rudnicki2018gauge,lin2019independent}. 
While gauge-invariant scoring metrics have been used in the past \cite{bgn+_robust_2013} (as we do in the present work), we note that the underlying representation of the gate set used to carry out GST is gauge-variant.
Specifically, GST eschews any notion of a fixed reference frame in favor of a \emph{gauge group} that specifies how to transform a valid estimate of an error model into a family of related error models which give identical experimental predictions.
While such gauge transformations do not impact the predictions made by such a model, they \emph{do} impact the particular channels reported at the end of the inference procedure, and some commonly reported metrics on channels are not gauge-invariant. 
In practice, one gauge-fixes resulting channels to some external reference frame, but this procedure requires nonlinear optimization whose global convergence is not guaranteed. 
Finally, a procedure for systematically including prior information in GST has not yet been put forward, which could potentially result in massive savings if developed.

In this paper, we introduce \emph{operational quantum tomography} (OQT), which is a general (operational) framework that allows us to reason about a host of different tomographic procedures (including GST) in a manifestly gauge-independent manner.
In addition to resolving the gauge problem, OQT allows us to naturally include prior 
information in GST within Bayesian inference, which was computationally prohibitive 
previously due to the the gauge fixing procedure.

OQT is enabled by using a new, manifestly gauge-invariant, representation of our gate set.  This representation is inspired by linear-inversion gate set tomography \cite{bgn+_robust_2013}; we term this the \emph{operational representation}.
After introducing the operational representation and explaining how it resolves the gauge problem, we discuss how to implement OQT numerically within a Bayesian framework while including prior information.
We then detail the performance of this technique by tracking prediction loss, a useful and gauge-invariant measure of the quality of our ability to predict the outcome of future experiments, across a suite of experimentally relevant problems: Ramsey interferometry, quantum process tomography, and randomized benchmarking.
We close by showing how dynamics of quantum systems may, in general, be described using the operational representation. 

\section{Gate set tomography and the operational representation}

\subsection{GST formalism}

As described in the introduction, quantum state and process tomography make strong assumptions about our ability to perform state preparation and measurement (SPAM). 
Tomographic reconstructions of states and processes that are made assuming perfect SPAM will be inconsistent with the true, noisy operations. 
A key advantage of GST is that it produces \emph{self-consistent} estimates by simultaneously characterizing SPAM along with other processes. 

\begin{figure}
    \begin{center}
        \includegraphics[width=\columnwidth]{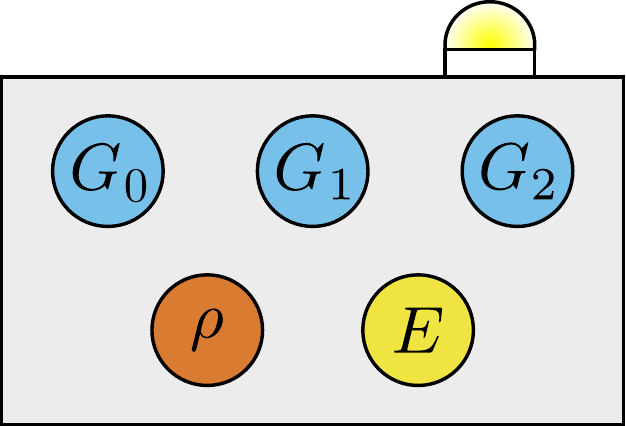}
    \end{center}
    \caption{
        \label{fig:black-box}
        An example of a quantum device modeled as a black box with buttons. Buttons are labeled by the actions they perform, for example prepare state $\rho$, apply operation $G_i$, and take measurement $E$. A light on top of the box turns on or stays off to indicate the result of the measurement. 
    }
\end{figure}

Here we briefly review GST, following 
Refs.~\cite{mgs+_selfconsistent_2013,bgn+_robust_2013,BlumeKohoutDemonstrationqubitoperations2017}, restricting our attention to the simplest case with a single state preparation and a single, two-outcome measurement.
To start, suppose that we have the ability to prepare an (unknown) state $\rho$, perform an (unknown) two-outcome measurement $E$, and perform some number $n$ additional (unknown) operations $\{G_0, \ldots G_{n-1}\}$.
We think about such a system as a box with labeled buttons, as depicted in \autoref{fig:black-box}, where each button denotes an operation we can perform. 
Hence, we have a button for state preparation ($b_{\rho}$), measurement ($b_E$), and buttons for each other operation
 labeled by elements of the set $\BS \defeq \{b_0, \dots, b_{n - 1}\}$, where we abbreviate $b_i = b_{G_i}$ for notational convenience.  A light on the box turns on or stays off to indicate the outcome of the measurement.
 
Within this formalism, all experiments we can perform are of the form:
\begin{enumerate}
    \item Press $b_{\rho}$ to begin the experiment.
    \item Sequentially press zero or more buttons from the set $\{b_0, \dots, b_{n - 1}\}$.
    \item Press $b_{E}$ to end the experiment.
    \item Record whether the light turned on.
\end{enumerate}

Our goal is to compute the likelihood of observing the light given a particular sequence of buttons. 
Within a quantum model, we do this by expressing the actions of buttons as \emph{super-operators}, which are linear operators that take density matrices to density matrices.
Formally, let $\Hil = \CC^d$ be a Hilbert space of finite dimension $d$.
Then, we denote by $\Lin(\Hil)$ the space of linear functions $\Hil \to \Hil$, and denote by $\Tra(\Hil)$ the space of linear functions $\Lin(\Hil) \to \Lin(\Hil)$.
Since $\Tra(\Hil)$ is a space of linear functions, elements of $\Tra(\Hil)$ can be written down as linear operators acting on vectors in $\Lin(\Hil)$. 
We denote vectors in $\Lin(\Hil)$ by ``super-kets'', e.g. $\sket{\rho}$; covectors for $\Lin(\Hil)$ are ``super-bras'' and correspond to measurements, e.g., $\langle\langle E|$.

As an example, if $d=2$, then $\rho$ can be represented as a $2\times2$ matrix, which we can instead arrange as a $4\times1$ column vector $\sket{\rho} \in \CC^{4}$. 
In this case, we can represent elements from $\Tra(\Hil)$ as $4\times4$ matrices, which act linearly (by multiplication) on super-kets.  \footnote{This is because, if $\text{vec}(A)$ denotes the vectorization of 
matrix $A$ (by column-stacking), then $\text{vec}(ABC)=(C^T\otimes A)\text{vec}{(B)}$.  As a density matrix $\rho$ evolves by similarity transformation (or a sum over them), $\sket{\rho}$ evolves by matrix multiplication.}
We assign to each button $b_i$ a super-operator $G_i \in \Tra(\Hil)$ (which we represent as a matrix acting on $\CC^{d^2}$). 
If $\Lambda_{G_i}$ is a quantum channel acting on a density matrix $\rho$, then $G_i$ is the operator such that $G_i \sket{\rho} = \sket{\Lambda_{G_i}[\rho]}$.  

We refer to the set of button presses between applications of SPAM as a \emph{sequence}. 
We denote the set of possible sequences as $\Seq$, which contains the empty sequence, 
as well as every possible combination of button presses. Sequences compose under 
concatenation.
\footnote{The set of 
experimental sequences $\Seq$ is a \emph{monoid} under addition; that is, $\Seq$ is 
closed under concatenation, and has the empty sequence of buttons $()$ as an additive 
identity.
We note that $\Seq$ does not contain inverses, since we cannot make a sequence of 
buttons shorter by pressing more buttons.} 
For two sequences $\vec{s}, \vec{t} \in \Seq$:
\begin{multline}
    (b_{s_0}, \dots, b_{s_{m - 1}}) + (b_{t_0}, \dots, b_{t_{m' - 1}}) = \\
    (b_{s_0}, \dots, b_{s_{m - 1}}, b_{t_0}, \dots, b_{t_{m' - 1}}).
\end{multline}
We will also write $|\vec{s}|$ to mean the length of $\vec{s}$, such that $|\vec{s} + \vec{t}| = |\vec{s}| + |\vec{t}|$.

Using the assignment of super-operators to buttons, we can compute the likelihood of experimental outcomes. 

\begin{definition}
 Let $\vec{s} = (b_{s_0}, b_{s_1}, \ldots b_{s_{m-1}})$ be a sequence of $m$ button presses from the buttons on our box.  
 The likelihood of the light turning on after performing $\mathbf{s}$, the \emph{sequence probability}, is given by the Born rule:
\begin{multline}
    \label{eq:example-gst-likelihood}
    \Pr(\textrm{light} | (b_{\rho}, b_{s_0}, \dots, b_{s_{m-1}}, b_E)) = \\
        \sbraket{E | G_{s_{m - 1}} G_{s_{m - 2}} \cdots G_{s_0} | \rho}.
\end{multline}
\end{definition}

\noindent This shows that, were we to learn the explicit form of $\sket{\rho}, \sbra{E}, \{G_i\}$, 
we would be able to predict the results of any future experiment. 
Nevertheless, as we already touched on in the introduction, super-operators suffer from a \emph{gauge} problem, making many numerically distinct sets of super-operators operationally equivalent.  (In the language of super-operators, $\{\sket{\rho}, \sbra{E}, \{G_i\}\}$ is gauge-equivalent to $\{B\sket{\rho}, \sbra{E}B^{-1}, \{BG_iB^{-1}\}\}$ for any appropriately-sized invertible matrix $B$.)
In the next section, we clarify this notion within the context of linear-inversion GST.

\subsection{Linear inversion GST}

The simplest GST inference procedure to learn $\sket{\rho}, \sbra{E}, \{G_i\}$ is linear-inversion GST (LGST). 
For any GST protocol, one first chooses a set of \emph{fiducial} sequences, $\vec{f} = \{\vec{f}_i\}$, which act as a ``reference frame'' for analysis of the experiments.
\footnote{It is sometimes necessary to pick distinct \emph{preparation} and \emph{measurement} fiducial sets; our results extend to those situations as well.}
Fiducial sequences are typically short sequences of button presses, and as a set they must be informationally complete (which will be formally defined below, with a consequence being that the set of fiducials has at least $d^2$ elements).

We use the set of fiducial sequences to construct the following scalar quantities:
    \begin{eqnarray}
        \tilde{E}_i &=& \sbra{E} F_i \sket{\rho}, \nonumber\\
        \tilde{F}_{ij} &=& \sbra{E} F_i F_j \sket{\rho}, \\
        \tilde{G}^{(k)}_{ij} &=& \sbra{E} F_i G_k F_j \sket{\rho}, \nonumber
        \label{eq:lgst}
    \end{eqnarray}

\noindent where $F_i$ is the super-operator obtained by multiplying together the super-operators for the constituent buttons in a fiducial sequence $\vec{f}_i$. 
In principle, the entries of these matrices are the probabilities of the light 
turning on for the given experiments. 
Hence, by repeating the experiments, we approximate these probabilities via the 
empirically observed frequencies.
Defining $A = \sum_j \sket{j} \sbra{E} F_j$ and $B = \sum_j F_j \sket{\rho} \sbra{j}$, 
where the $\sket{j}$ are basis states of the space $\Hil \otimes \Hil$, we can recover the desired $\sket{\rho}, \sbra{E}$, and $\{G_k\}$ according to:
\begin{eqnarray} \label{eqn:LGST_step}
\sket{\rho} &=& B\tilde{F}^{-1}\tilde{E} \nonumber\\
\sbra{E} &=& \tilde{E}^T B^{-1} \\
G_k &=& B \tilde{F}^{-1} \tilde{G}^{(k)}B^{-1} \nonumber 
\end{eqnarray}
We further note that, by definition, $\tilde{F} = AB$.  
We require that $\tilde{F}$ has rank of at least $d^2$, where $d$ is the dimension of
the qubit Hilbert space. 
If $\mathrm{dim}(\tilde F) > d^2$, then the pseudo-inverse is used instead of the
normal inverse.  
This rank criterion ($\mathrm{rank}(\tilde{F})=d^2$) is what provides our definition of informational completeness in this context.
It provides a check of the choice of fiducials, which can be useful if good initial guesses of the gates are not known.
Such a set of fiducials can even be chosen  `on the fly' by performing experiments until one can construct an invertible $\tilde{F}$. 

\subsection{Gauge and the operational representation}

In the above section, one might be troubled that we did not actually recover the literal values $G_k, \rho,$ and $E$.
Rather, they are now complicated by the presence of the gauge transformation $B$ - the ``true'' super-operators $G_k, \rho, E$ are all gauge-dependent quantities.
However, the gauge $B$ itself is not accessible experimentally. 
This is because the observed sequence probabilities are totally independent of gauge:
\begin{multline}
    \label{eq:gauge-independence}
    \Tr \left( \sket{\rho} \sbra{E} G_{s_{m - 1}}  \cdots G_{s_0} \right) = \\ \hbox{Tr} \left( B^{-1} \sket{\rho} \sbra{E} B B^{-1}  G_{s_{m - 1}} B \cdots B^{-1}  G_{s_0} B \right) 
\end{multline}

More formally, let us begin by making a mapping between button sequences and super-operators.
We assign an element of $\Tra(\Hil)$, the space of linear operators on $\Hil$, to each sequence $\vec{s} \in \Seq$ using a mapping $\Phi : \Seq \to \Tra(\Hil)$,
\begin{align}
    \Phi((b_{s_0}, \dots, b_{s_{m - 1}})) = G_{\vec{s}}.
\end{align}
In general, the mapping $\Phi$ between button sequences and channels can arbitrary, especially in the presence of non-Markovian errors, but in this work we will consider the special case in which $\Phi$ is a homomorphism between the monoids $\Seq$ and $\Tra(\Hil)$ \footnote{Note that $\Tra(\Hil)$ is monoid under multiplication rather than concatenation. In general, $\Tra(\Hil)$ fails to be a group, as we cannot invert general quantum operations due to decoherence. (Decohering channels are invertible, as long as they are full rank, but these inverses are not completely positive, and are therefore unphysical.)  We can then view $\Phi$ as a homomorphism from button sequences to super-operators, since $\Phi(s + t) = \Phi(t) \Phi(s) \text{ and } \Phi(()) = \id$.
Since we have listed sequences left-to-right rather than right-to-left, $\Phi$ is formally a homomorphism from button sequences to the opposite monoid of super-operators, defined by the opposite product $A {} {\cdot}^{\op} B \defeq BA$, but we ignore this detail as a notational convenience.},
\begin{align}
    \Phi((b_{s_0}, \dots, b_{s_{m - 1}})) \defeq G_{s_{m - 1}} \cdots G_{s_0}.
\end{align}
This mapping is not unique, but can be specified by the outputs of $\Phi$ for each single-button sequence, $\Phi((b_0)) = G_{b_0}$, $\Phi((b_1)) = G_{b_1}$, and so forth.
Considering this special case, we can think of SPAM as a special button $b_{\text{SPAM}}$ such that
\begin{align}
    \Phi((b_{\text{SPAM}})) = \sket{\rho} \sbra{E}.
\end{align}
Making this identification, we can then use $\Phi$ to recover the probabilities in \autoref{eq:example-gst-likelihood} by taking the trace of $\Phi(\vec{s})$ for each sequence $\vec{s} \in \Seq$, so long as we adopt the convention that $\vec{s}$ begins with $b_{\text{SPAM}}$,
\begin{align}
    \Pr(\text{light} | \vec{s}; \Phi) = \Tr(\Phi(\vec{s})).
\end{align}

The problem of inferring the properties of our box is equivalent to identifying which $\Phi$ maps from button sequences to super-operators in a manner that correctly predicts experimental outcomes according to \autoref{eq:example-gst-likelihood}.
Following this motivation, we define that two mappings $\Phi, \Phi' : \Seq \to \Tra(\Hil)$ are \emph{gauge-equivalent} ($\Phi \sim \Phi'$) if and only if they yield the same sequence probabilities for all elements of $\Seq$. The term ``gauge'' used to describe the equivalence class $\sim$ is motivated by the observation that
\begin{multline}
    \Phi \sim \Phi' \text{ if and only if there exists } B \in \GL(\CC^{d^2}) \\ \text{ such  that for all } \vec{s} \in \Seq: 
    \Phi(\vec{s}) = B\Phi'(\vec{s})B^{-1}.
\end{multline}

We say that the equivalence class $[\Phi] \defeq \{\Phi' \in \Hom(\Seq, \Tra(\Hil)) \text{ such that } \Phi' \sim \Phi\}$ of gauge-equivalent $\Phi$ is the \emph{gauge orbit}. 
It is easy to identify one such $\Phi$ (just choose any invertible matrix of appropriate dimension), but it is expensive to compute an entire equivalence class of distinct ones.

\begin{figure*}
    \begin{center}
        \includegraphics[width=\textwidth]{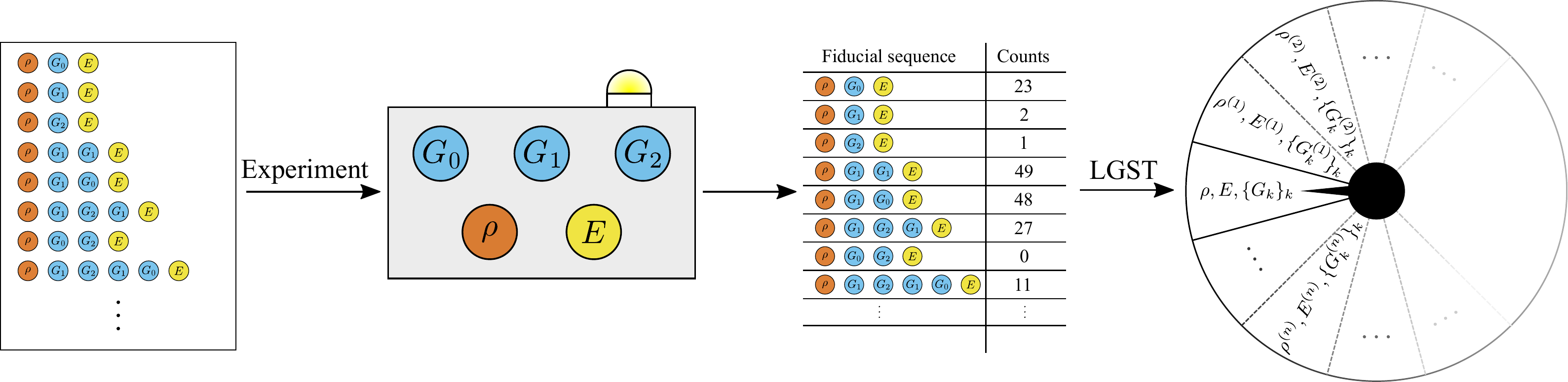}
    \end{center}
    \caption{
        \label{fig:gauge-orbit}
      Pipeline for linear inversion gate set tomography (LGST). A set of fiducial sequences is chosen; we perform the specified experiments and record how many times the light turned on. Following the linear inversion step in \autoref{eqn:LGST_step}, we can reconstruct a copy of the super-operators for each button. However, the results we obtain will be expressed in an unknown gauge which is one of infinitely many in the gauge orbit.
    }
\end{figure*}

Choosing a gauge to represent a gate set is typically accomplished through nonlinear optimization, in which 
a gauge is sought that transforms the estimated gate set to be as close as possible (by some metric) to an ideal ``target'' gate set.
This allows for computation of gauge-variant metrics between the estimate and the target (e.g., diamond distance, fidelity).
In practice, these procedures can work reasonably well \cite{BlumeKohoutDemonstrationqubitoperations2017}, 
but they scale inefficiently, are not guaranteed to be numerically stable, and are not guaranteed to not get stuck in a local extremum.
Thus, as a practical matter, we would like to identify a set of parameters that is necessary and sufficient to identify gauge orbits without having
to actually perform a gauge optimization.
That is, we seek to parameterize and perform inference on the set of gauge orbits directly:
\begin{align}
    \Gua(\BS, \Hil) & \defeq \Hom(\Seq, \Tra(\Hil)) / {\sim} = \{[\Phi]\},
\end{align}
where $A / {\sim}$ is the factor set of $A$ defined by the relation $\sim$ as the set of equivalence classes $A / {\sim} \defeq \{[a] : a \in A\}$.

When it is clear from context which button set and Hilbert space are used to define our box, we will omit them for brevity, writing that $\Gua = \Gua(\BS, \Hil)$.
We say that each member of $\Gua(\BS, \Hil)$ is a \emph{gate set}, such that identifying which member of $\Gua(\BS, \Hil)$ was used to generate a data record is \emph{gate set tomography}.
When it is clear from context, we will also refer to sets of super-operators $\GS = \{G_0, \dots, G_{k - 1}, \sket{\rho}, \sbra{E}\}$ as gate sets, with the implicit understanding that we are interested in the gauge orbit $[\GS]$ (equivalence class under $\sim$) of $\GS$.

We call any such representation of $\Gua(\BS, \Hil)$ \emph{operational}, since it is a complete description of all operational experiments that we can perform on our box, under the promise that the box is described by some model over $\Hil$.
In fact, we have already seen an especially convenient operational representation: $\tilde{E}, \tilde{F}, \{\tilde{G}^{(k)}\}$.
They are a set of gauge-independent values (as they are directly experimentally observable), and are unique to a particular gate set for a given choice of fiducials. 
They also yield the same measurement probabilities as their gauge-dependent counterparts. 
To see this, consider some sequence of button presses $(b_\rho, b_{s_0}, \ldots b_{s_{m-1}}, b_E)$.
 The sequence probability is:
 \begin{equation}
   \begin{split}
    & \Pr(\text{light} | (b_{\rho}, b_{s_0}, \dots, b_{s_{m - 1}}, b_E)) \\ 
    &= \sbraket{E | G_{s_{m - 1}} \cdots G_{s_0} | \rho} \nonumber \\
    &= \hbox{Tr} \left( \sket{\rho} \sbra{E} G_{s_{m - 1}} \cdots G_{s_0} \right) \nonumber  \\
    &=  \hbox{Tr} \left( B^{-1} \sket{\rho} \sbra{E} B B^{-1}  G_{s_{m - 1}} B  \cdots B^{-1}  G_{s_0} B \right) \nonumber  \\
    &= \hbox{Tr} \left( \tilde{F}^{-1} \tilde{E} \tilde{E}^\Tra  \tilde{F}^{-1} \tilde{G}^{(s_{m-1})}  \cdots \tilde{F}^{-1} \tilde{G}^{(s_0)} \right).
   \end{split}
 \end{equation}
This leads to the remarkable fact that \emph{when we learn $\tilde{E}, \tilde{F}, \{\tilde{G}^{(k)}\}$, we can predict the outcome of any future experiments.}
Note that this statement is distinct from performing LGST: we can use $\tilde{E}, \tilde{F}, \{\tilde{G}^{(k)}\}$ as our underlying model, while updating it via more sophisticated experiments.

\section{Implementation}

Having thus established that learning the operational representation of a gate set allows us to predict its behavior, we are left with the question of \emph{how} to learn operational representations from data records.
In this section, we describe our implementation of operational quantum tomography, based on Bayesian inference.
In particular, we implement the inference numerically using the particle filter, or sequential Monte Carlo (SMC) approach, a standard technique for carrying out Bayesian inference computationally \cite{dj_tutorial_2011}.

\subsection{Bayesian inference: obtaining posteriors from evidence}

As applied to quantum information, Bayesian inference is a formalism for describing our knowledge about a quantum system given classical data observed from it.
In particular, Bayesian inference represents our state of knowledge at any given point in a characterization protocol by a distribution of the form $\Pr(\text{hypothesis} | \text{data})$, where ``$\text{hypothesis}$'' describes some hypothesis that we can use to predict the future behavior of our quantum system, and ``$\text{data}$'' is the set of observations made of that system.

In the special case that $\text{data} = \{\}$ (that is, before we have made any observations), we write our state of knowledge as $\Pr(\text{hypothesis})$, also known as our \emph{prior distribution}.
For example, in traditional Ramsey interferometry, our hypothesis might consist of the assumption that the system evolves under a Hamiltonian of the form $H = \omega \sigma_z / 2$ for some $\omega$. We may assign a prior distribution over $\omega$ such as 
\begin{align}
    \Pr(\omega) = \begin{cases}
        1 / {\omega_{\max}} & \omega \in [0, {\omega_{\max}}] \\
        0 & \text{otherwise},
    \end{cases}
\end{align}
representing that we are equally willing to believe that $\omega$ has any value in the interval $[0, {\omega_{\max}}]$.

Since distributions of the form $\Pr(\text{hypothesis} | \text{data})$ represent our state of knowledge at any point during an experimental procedure, equipped with such a distribution, we can answer questions such as ``what is the best hypothesis to report given what we have learned from our quantum system?''.
Returning to the Ramsey example, we may want to report an estimate $\hat{\omega}$ such that the the squared error $(\hat{\omega} - \omega)^2$ is minimized on average.
As summarized in Appendix~\ref{apx:review-bayes-est}, this is achieved by reporting the Bayesian mean estimate $\hat{\omega}_{\BME} \defeq \expect_{\omega}[\omega | \text{data}] = \int \omega \Pr(\omega | \text{data}) \dd\omega$.

We are thus left with the problem of finding our state of knowledge at some point in an experimental procedure given our most recent observation, and given our previous state of knowledge; that is, of how to update our state of knowledge to reflect new information.
To do so, we rely on Bayes' rule, which states that
\begin{multline}
    \label{eq:bayes-rule}
    \Pr( \hbox{hypothesis} | \hbox{data}) \sim \\ 
    \Pr (\hbox{data} | \hbox{hypothesis}) \times  \Pr (\hbox{hypothesis}),
\end{multline}
where $\Pr (\hbox{hypothesis})$ is our prior distribution, and $\sim$ indicates equality up to renormalization.
Intuitively, this rule tells us that a hypothesis is reweighted according to how plausible it is for a given observation to arise given that hypothesis.
To perform this update, we must simulate $\Pr(\text{data} | \text{hypothesis})$, known as the \emph{likelihood function} for our quantum system.
Put differently, we can only learn properties of a system whose effects can be simulated.
We cannot learn about a parameter that has no effect on the outcomes of system, or whose effects we cannot simulate.

It is for this reason that, in the rest of the paper, we take our hypothesis to be the operational representation of some quantum system.
In particular, the operational representation is a minimal set of parameters required to simulate the behavior of that system, such that any parameter beyond the operational representation cannot have any effect on our predictions.
For example, we can never learn gauge parameters from experimental observations, as they have no effect on the likelihood function for any measurement that we could perform \footnote{This argument shows that the use of operational representations can be motivated by appeal to the \emph{likelihood principle}, which informally states that all inference --- whether or not carried out using Bayesian reasoning --- must depend on a system only through its likelihood function.}.

\subsection{Numerical approach: sequential Monte Carlo}
\label{subsec:particle-filter}

So far, we have regarded Bayesian inference in the abstract, without reference to or concern for how one might implement an inference procedure in practice.
A practitioner interested in using Bayes' rule will find it difficult to work with \autoref{eq:bayes-rule} directly, as the normalization suppressed by the use of $\sim$ notation converges exponentially quickly to 0 with the amount of data considered, exacerbating numerical precision issues.
Moreover, any choice of discretization informed by the prior is not likely to be terribly useful as the posterior shrinks in width.

In lieu of these considerations, a number of different computational algorithms have been developed that offer a Bayesian practitioner a range of different options.
For instance, rejection sampling techniques such as the Metropolis--Hastings 
algorithm \cite{hastings1970}, as well as more sophisticated modern algorithms such as Hamiltonian Monte Carlo \cite{bet_conceptual_2017}~and NUTS \cite{hg_nouturn_2011}, allow for obtaining samples from a posterior distribution with reasonable computational effort.
These algorithms have been used in quantum information to solve otherwise intractable 
problems such as the estimation of randomized benchmarking parameters \cite{HincksBayesianInferenceRandomized2018}.

For application to online experimental protocols, however, it is often useful to adopt an algorithm that works in a \emph{streaming} fashion. 
This allows for samples from a posterior distribution to be drawn at any point in an experimental procedure, such that adaptive decisions such as stopping criteria or experiment design can be made easily. 
Critical to realizing this capability is that the cost of an algorithm can depend only approximately linearly on the amount of data taken. 
This restriction motivates the use of filtering algorithms, which update an approximation of a prior given incoming data to yield a new approximation of the resulting posterior.
The Kalman filter, for example, is a Bayesian filter for the special case in which the prior and posterior are both normal, and in which the likelihood is a linear model perturbed by normally distributed noise \cite{bbm_bayesian_1995}.

In this paper, we adopt the \emph{particle filter} \cite{dj_tutorial_2011}, also known as the sequential Monte Carlo approximation. 
Particle filters are applicable to a very broad range of likelihood functions, and give rich diagnostic data to assist in understanding their execution. 
The QInfer library \cite{gfh+_qinfer_2016}~provides a useful implementation of particle filters for quantum information applications, and this library is used throughout the rest of the work.

Particle filters work by representing the distribution over some random variable \(\vec{x}\) as a weighted sum of \(\delta\)-distributions at each step,
\begin{align}
    \Pr(\vec{x}) \approx \sum_i w_i \delta(\vec{x} - \vec{x}_i),
\end{align}
where $\{w_i\}$ are non-negative real numbers summing to 1, and where $\{\vec{x}_i\}$ are different hypotheses about $\vec{x}$. 
Each hypothesis $\vec{x}_i$ is called a \emph{particle}, and is said to have a corresponding weight $w_i$. Numerical stability is achieved by periodically moving each particle to concentrate discretization on regions of high posterior density \cite{lw_combined_2001}.
Examples of this in operation can be seen in videos at \url{https://youtu.be/aUkBa1zMKv4}~and \url{https://youtu.be/4EiD8JcCSlQ}.

\subsection{Setting priors over the operational representation}\label{sec:setting_priors}

Within a Bayesian framework, we begin with a statement about our beliefs before starting an experiment.
We write this down formally as a \emph{prior distribution}, which gives us a mathematical description of our prior knowledge.
In absence of any data from a particular experimental run, a prior distribution $\pi$ assigns a probability $\pi(\vec{x})$ to each object of interest $\vec{x}$ (\emph{e.g.,} the elements of the operational representation).

In experimental QCVV, we typically express our beliefs in terms of gauge-dependent formalisms (\emph{e.g.}, super-operators).
Here, we need to translate these prior distributions into a prior over the operational representation, which is gauge-independent.
Fortunately, we can easily sample from the prior distribution over the operational representation induced by a distribution over a gauge-dependent representation. Upon choosing a set of fiducial sequences, we proceed to:
\begin{enumerate}
\item State the prior over some gauge-dependent representation (\emph{e.g.}, parameters in super-operators).
\item Draw a sample from the gauge-dependent prior.
\item Convert the gauge-dependent sample to the operational representation by applying LGST. 
\item Return this as the sample from the gauge-independent prior.
\end{enumerate}

As a concrete example, suppose we intuit that a particular button should perform single-qubit $Z$-rotation gates.
We can write these in a familiar, gauge-dependent way by expressing them as super-operators in some matrix basis, $R_z(\theta)$ (in our implementation, we use the Pauli basis).
Now suppose that we suspect this button over-rotates about $Z$ by an angle $\delta \theta$ that is somewhere between $0$ and $\pi / 10$. 
To generate samples from this prior expressed in the operational representation according to this belief, we first choose samples of $\delta \theta$ uniformly at random from $0$ to $\pi / 10$. 
Next, we use these sampled angles to synthesize corresponding channels for each member of the gate set, i.e. $R_z(\theta + \delta \theta)$.
A prior distribution in terms of superoperators can be constructed in a similar manner for each button on the box.
Together, we use them to compute the frequencies for each element of the operational representation using the linear-inversion step of \autoref{eq:lgst}.

\subsection{Informational completeness and germ sensitivity}

In addition to choosing a prior distribution, we must also choose a set of fiducial sequences to fix a reference frame. 
Any choice will yield a valid operational representation, in the sense that we can populate an $\tilde{E}, \tilde{F}$, and $\{ \tilde{G}^{(k)} \}$ with the outcome frequencies of the experiments. 
However, an additional requirement of the fiducial sequences is that they must be \emph{informationally complete}. 
As we will see, the definition of this is dynamic.

Consider for a moment standard quantum state tomography, where we prepare (perfectly) some unknown state, and can execute perfect measurements. 
In the case of a single qubit, it is well known that measuring $\sigma_x$, $\sigma_y$, and $\sigma_z$ is sufficient to fully reconstruct the state \cite{nielsen2002}.
These measurements span the Bloch sphere, and we say that they are informationally complete.

In GST, a similar notion holds. 
However, we do not know \emph{a priori} how the measurement and operation buttons are oriented relative to an external reference frame.
For instance, if someone provides us with a box with buttons labeled $\sigma_x$, $\sigma_y$, $\sigma_z$, we do not know what they actually do. 
They may be noisy implementations of these operations, they may be completely different operations, or, they may even do nothing at all. 
Naively using these buttons to execute measurements is therefore not guaranteed to give us something informationally complete, even if the labels suggest they should.

In GST, we can check for informational completeness using the matrix $\tilde{F}$ in the operational representation. 
Recall that this is constructed using experiments performed by applying pairs of the fiducial sequences. 
When we initialize the operational representation from the prior over super-operators, we must compute $\tilde{F}^{-1}$. 
If the fiducial sequences are poorly chosen, $\tilde{F}$ may be ill-conditioned, or even singular. 

\begin{definition}
    \label{def:info-complete}
    A set of fiducial sequences $\vec{f} \subset \Seq$ is 
    \emph{informationally complete} for a gate set $\GS$ if $\tilde{F} = \sum_{ij} 
    \sbra{E} F_i F_j \sket{\rho}\sket{i}\sbra{j}$ has rank of at least $d^2$, where $d$ is the dimension
    of the qubit Hilbert space. Here, $F_i = \Phi[\vec{f}_i]$ and $\sket{\rho}\sbra{E} = \Phi[\SPAM]$ 
    for some $\Phi \in \GS$.
\end{definition}

Note that since we can conjugate by $B$ in Definition~\autoref{def:info-complete}, we can choose any $\Phi \in \GS$ that is convenient for evaluating $\tilde{F}$ --- if we can find a $\Phi$ such that a set of fiducial sequences is informationally complete, it must also be complete for all $\Phi' \in [\Phi]$.

An issue that arises as a consequence of the choice of fiducials is that it is possible to find values across $\tilde{E}, \tilde{F}, \{\tilde{G}^{(k)}\}$ that are identical.
For example, if one of the fiducials is the empty sequence, then $\tilde{E}_0 = \tilde{F}_{00}$, $\tilde{E}_1 = \tilde{F}_{01} = \tilde{F}_{10}$, and so forth. 
Were we to perform a SMC update step on the full set of matrix entries, the entries that are constrained to be identical will be perturbed in different ways, leading to inconsistent outcomes.

To remedy this, we first perform a preprocessing step that eliminates redundant entries, producing
a minimal set of model parameters on which we can perform inference. 
Mappings are employed to transform the minimal set back to full $\tilde{E}, \tilde{F}, \{\tilde{G}^{(k)}\}$, and \emph{vice versa}, throughout. 
Learning this minimal set of parameters is then sufficient to characterize the entire system.
This trimming procedure also has the benefit of substantially reducing the number of model parameters required, speeding up the inference process.

Beyond fiducial selection, we have considerable freedom in the selection of the particular experiments we perform. 
The best choice of experiments depends on our particular learning objective. 
For most of our demonstrations in this work, we fix a total number of experiments, a minimum/maximum sequence length, and then produce sequences of increasing lengths between the bounds. 
In some implementations of GST, one designs a small collection of short button sequences known as ``germs", such that by taking appropriate powers of germs one can amplify coherent errors to gain optimal information as the number of experiments is increased.
Such a pattern of germs is called ``amplificationally complete" 
\cite{BlumeKohoutDemonstrationqubitoperations2017}, and can reduce the total number
of experiments required. 
We take this approach in our implementation of GST, and take care to identify the experiments we carry out in all of our examples.

\subsection{Constraints on gates and gate sets}

If we represent inferred channels with super-operators (as is customary in quantum process tomography), the allowed form of the matrices is not arbitrary.
Rather, physical constraints such as positivity and trace preservation of density matrices restrict the allowed structure.
When generalizing to GST, the problem of identifying when a gate set is valid is complicated by the introduction of the gauge. 
The elements of a gate set might not, in a particular representation, be CPTP, but are gauge equivalent to a CPTP representation.
Performing inference on an operational representation introduces a similar challenge: how do we ensure that an operational representation corresponds to a gate set that makes physical sense? 

Analogous to the case in GST, we need a condition that is simultaneously satisfied by all the gates in the gate set.
For the operational representation, an obvious first test is to check whether all the entries are in the interval $[0,1]$. 
Since entries in the operational representation correspond to sequence probabilities, this is a necessary physical constraint.
This is not a sufficient condition, though, as the probabilities for any possible future experiment must be constrained in the same way. 
This leads us to the following definition:

\begin{definition}[Positivity]
An estimate of a gate set $\hat{\mathcal{G}}=\{\hat{G}_0,\ldots, 
\hat{G}_{n-1}, \sket{\hat \rho},\sbra{\hat E}\}$ is \emph{positive} 
if for all $\hat{\mathcal{S}} \in \{\hat{G}_0,\ldots,\hat{G}_{n-1}\}^\star$, where 
$\{\cdot\}^\star$ is the Kleene-closure\footnote{The Kleene closure $S^\star$ of a 
set $S = \{s_0, s_1, \dots\}$ is given by the set of all finite-length strings over 
$S$, $S^\star = \{(), (s_0), (s_1), \dots, (s_0, s_0), (s_0, s_1), \dots, (s_1, s_0), \dots, 
$ $(s_0, s_0, s_0), \dots\}$.}, we have that both $\sbra{\hat{E}} \hat{\mathcal{S}} 
\sket{\hat{\rho}}\ge 0$ and $\sbra{\openone - \hat{E}} \hat{\mathcal{S}} 
\sket{\hat{\rho}}\ge 0$.
\end{definition}
   
Other than by converting to the operational representation a standard (gauge-variant) 
representation which is explicitly positive (in some gauge),
it is unclear how one can create operational representations that are positive by construction. 
However, it \emph{is} possible to ensure that inference begins from a point where this 
is true through our choice of prior distribution.
As described in \autoref{sec:setting_priors}, when we set a prior distribution, we 
begin with a gauge-dependent prior.
When we do this, we express each gate of the gate set in the same gauge that we have chosen.
We can then guarantee by construction that each member of the gate set has 
characteristics such as complete positivity, to ensure they will always produce valid 
outcome probabilities.

As inference proceeds, however, checking for properties such as complete positivity is
practically difficult. 
This is because to check such properties, one needs to perform a gauge-fixing procedure,
which we wish to avoid for aforementioned reasons.
Such a procedure is not impractical to do once at the end of an inference procedure, but
it is at each update step during Bayesian inference. 

One workaround to this is to ensure  at the very least all the values in the operational 
representation are positive. 
Though this of course doesn't guarantee true positivity, we found in practice that 
negative values of the likelihood function appear regularly, and these must be handled 
appropriately in order for the sequential Monte Carlo updates to succeed. 
As a workaround, we simply clip the output of the likelihood function so that any
negative `likelihoods' are set to 0, and any positive likelihoods greater than $1$ are set to 1.

An alternate way to approach model validity is to choose a set of validation experiments. 
In plausible experimental settings, one has a specific application (and hence gate sequence) in mind.
We can then decide if a model is valid for a particular set of gate sequences by checking if it produces a proper
likelihood for all the validation experiments, a notion that we call \emph{operational positivity}.

\begin{definition}[Operational positivity]\label{eq:op_pos}
    An estimate of a gate set $\hat{\mathcal{G}}=\{\hat{G}_0,\ldots, 
    \hat{G}_{n-1}, \sket{\hat \rho},\sbra{\hat E}\}$ is \emph{operationally positive} on a 
    set $\hat{\mathcal{S}}_{\rm test} \subseteq \{\hat{G}_0,\ldots,\hat{G}_{n-1}\}^\star$ if for all $\hat{\mathcal{U}}\in\hat{\mathcal{S}}_{\rm test} $  both $\sbra{\hat{E}}\hat{\mathcal{U}} \sket{\hat{\rho}}\ge 0$ and $\sbra{\openone - \hat{E}} \hat{\mathcal{U}} \sket{\hat{\rho}}\ge 0$.
\end{definition}

From these definitions, positivity implies operational positivity but the converse need not be true.
Operational positivity is a useful concept because it is both easy to test and also of practical relevance when one wishes to check particular applications.
In our work, we make extensive use of operational positivity, since whenever the sequential Monte Carlo inference procedure resamples particles, it is necessary to avoid negative predicted probabilities.

\section{Prediction Loss}
\label{sec:predloss}

Once we have obtained a posterior distribution $\Pr(\vec{x} | \text{data})$ over the operational representation $\vec{x}$ of our gate set from some sequence of experiments, we are typically interested in extracting diagnostic and benchmarking information.
To do so in a manner consistent with the gauge, one could 
consider \emph{gauge fixing} procedures, which consist of optimization problems that 
pick out particular gauge-dependent representations of a gate set that we can then use 
to report traditional metrics \cite{BlumeKohoutDemonstrationqubitoperations2017}.

For instance, if we intend \emph{a priori} that the $b_x, b_y, b_z$ buttons should be describable by unitary transformations $\e^{-\ii \pi \sigma_x / 2}$, $\e^{-\ii \pi \sigma_y / 2}$, and $\e^{-\ii \pi \sigma_z / 2}$, respectively, we may wish to report gauge-dependent metrics such as the diamond norm by fixing to a gauge that best agrees with this description.
By taking the best case over members of the gate set in that gauge we can construct statements such as ``there exists a gauge-dependent description of our gates such that with posterior probability at least $(1 - \alpha)$, the agreement between each gate and their action in a particular chosen frame is no worse than $\epsilon$.''

Unfortunately, this gauge-fixing procedure can be cumbersome to implement (especially across many hypotheses), is not guaranteed to work (i.e., find the optimum gauge given a target gate set) and is open to multiple interpretations.
As an alternative, we instead will score our predictions on a set of experiments of interest. 
To do this, we recall that a gate set $\GS$ is sufficient to predict the outcome of \emph{any} hypothetical experiment we may wish to perform within the GST framework by \autoref{eq:op_pos}.
We thus take a data-driven approach to the problem and choose a set of button sequences $S_{\validate}$ that we are interested in correctly predicting.
Concretely, let $p_{\vec{s}} \defeq \Pr(\hbox{light} | \vec{s})$ for each $\vec{s} \in S_{\validate}$ be a parameter that we are interested in estimating.
If we predict $\hat{p}_{\vec{s}}$ for $p_{\vec{s}}$, then we can consider the quadratic loss
\begin{equation}
    L_{\vec{s}}(\hat{p}_{\vec{s}}, p_{\vec{s}}) = \left( \hat{p}_{\vec{s}} - p_{\vec{s}}  \right)^2.
\end{equation}
We call this a \emph{prediction loss} for the sequence $\vec{s}$, since it rewards estimators that can accurately predict the outcome of future experiments. The quadratic loss is by no means unique, and there are other suitable choices, such as the Kullback-Liebler divergence.

Since each prediction loss function is Bregman for each $\vec{s}$, the Bayesian mean estimator (BME), where we average over the prediction made for each gate set in the support of our posterior, is optimal \cite{bgw_optimality_2005} \footnote{For a more detailed discussion of the role of Bregman estimators in tomography, we refer the reader to the work of \citet{fk_have_2015}.}.
That is, to minimize loss we choose as our estimator
\begin{align}
    \label{eq:prediction-bme}
    \hat{p}_{\vec{s}} = \expect_{\GS}\left[\Pr(\hbox{light} | \GS; \vec{s}) | \text{data} \right].
\end{align}
Intuitively, we predict the outcome of measuring the sequence $\vec{s}$ for each hypothesis $\GS$, and then take the average.

This gives us much better predictive capability than restricting ourselves to using a single estimated gate set to predict all future experiments.
As we validate with longer and longer sequences than those in the training set that we used to obtain our posterior in the first place, our posterior uncertainty in $p_{\vec{s}}$ will necessarily grow, as can be seen from the method of hyperparameters \cite{Granade_rohl_2012}.
This is not reflected if we pick a single gate set, but is immediately included in the Bayesian mean estimator \autoref{eq:prediction-bme}, which will tend to hedge towards $\nicefrac12$ as sequences grow in length.


\section{Examples}

In this section, we demonstrate the versatility of our framework by applying it to many common QCVV protocols. 
This includes replicating the results of other state-of-the-art techniques, such as long-sequence gate set tomography \cite{BlumeKohoutDemonstrationqubitoperations2017} and randomized benchmarking. 
A discussion of applications of OQT to quantum state tomography can be found in Appendix~\ref{app:statetomo}.

\begin{figure*}[tb]
    \centering
    \includegraphics[width=0.8\textwidth]{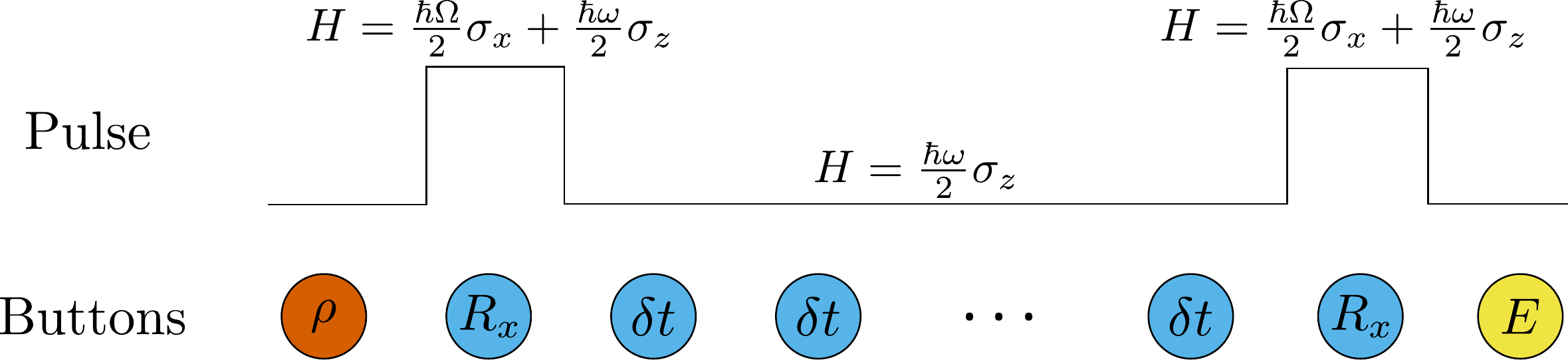}
    \caption{Depiction of Ramsey interferometry as a pulse diagram, and as a button sequence in the OQT framework.}
    \label{fig:pulsediagram}
\end{figure*}

-----------------------------
\subsection{Ramsey interferometry}
\label{subsec:ramsey}

Single-qubit operations are often implemented by applying electromagnetic pulses to induce rotations about the Bloch sphere. 
The basis of such methods is the intrinsic Rabi oscillation frequency of the system, which tells us the likelihood of measuring the qubit in its $\ket{0}$ or $\ket{1}$ state at a given time. 
Knowledge of the Rabi frequency allows us to adjust the pulse duration in order to obtain exactly the superpositio nwe desire.

\begin{table*}
    \centering
        \begin{tabular}{p{0.2\textwidth}p{0.5\textwidth}p{0.2\textwidth}}
            \hline \hline
            \textbf{Button label} & \textbf{Prior} & \textbf{Example values} \\
            $\rho$ & $1/\sqrt{2} \begin{pmatrix} 1 & 0 & 0 & 1 \end{pmatrix}$, depolarized with $p \in \mathcal{U}(0, 0.1)$ & $p = 0.038311$ \\
            \texttt{$R_x$} & $R_x\left(\pi/2 + \epsilon\right), \quad \epsilon \in \mathcal{N}(0, 10^{-3}) $ & $\epsilon = -0.003824$ \\
            $\delta t$ & $R_z(\omega \cdot \delta t), \quad \omega \in \mathcal{U}(0, 1) $ & $\omega  = 0.346754, \enskip \delta t = 1$ \\
            $E$ & $1/\sqrt{2} \begin{pmatrix} 1 & 0 & 0 & 1 \end{pmatrix}$, depolarized with $p \in \mathcal{U}(0, 0.1)$ &  $p = 0.023933$ \\ \hline \hline
        \textbf{Fiducial seqs.} & $\{(\cdot), \quad (R_x), \quad (R_x, \enskip R_x), \quad (R_x, \enskip \delta t, \enskip R_x)\}$ & \\
     \textbf{Training exps.} & $(R_x, \enskip (\delta t)^n, \enskip R_x)$ & $n = 2, \ldots, 49 $ \\
    \textbf{Testing exps.} & $(R_x, \enskip (\delta t)^n, \enskip R_x)$ & $n = 50, \ldots, 100 $ \\
    \hline \hline
        \end{tabular}
    \caption{
        \label{tab:ramsey_oqt}
        OQT parameter specification for Ramsey interferometry. Example values correspond to those used in the plots and provided example notebook. Button sequences are represented as lists, where the buttons are applied from left to right, and application of SPAM is implicit in the training and testing experiments. Button labels are abbreviated as $b_{R_x} \rightarrow R_x$ for notational simplicity.
    }
\end{table*}

Typically, one learns this frequency by means of either Rabi or Ramsey interferometry.
In Ramsey interferometry (depicted in \autoref{fig:pulsediagram}), a qubit is prepared in state $\ket{0}$ and then a $R_x\left(\frac{\pi}{2}\right)$ pulse is applied.
The qubit is left to evolve under the Hamiltonian $H = \frac{\omega t}{2} \sigma_z$ for some time $t$, after which another $R_x\left(\frac{\pi}{2}\right)$ pulse is applied, followed by measurement in the computational basis. 
The likelihood of obtaining the measurement outcome $\ket{0}$ is given by 
\begin{eqnarray}
    \Pr(0 | \omega; t) &=& |\bra{0}  R_x \left(\frac{\pi}{2} \right) 
                       e^{-i \omega t \sigma_z / 2} R_x \left(\frac{\pi}{2} \right) | 
                       0 \rangle |^2  \nonumber \\
                       &=& \cos^2 \left( \frac{\omega t}{2} \right)
\end{eqnarray}
However, in a given implementation, it is likely that the $R_x(\frac{\pi}{2})$ pulse is not perfect and the resultant state will be slightly over- or under-rotated, yielding an incorrect estimate of $\omega$. 
Our goal is to learn not only $\omega$, but also the precise rotation angle so that we can compensate for this discrepancy by adjusting the duration of the pulse.

First, we translate Ramsey interferometry into the operational framework language (a box with buttons).
In this case, there are four buttons.
The first two are for SPAM - the button $b_{\rho}$ prepares the $\ket{0}$ state, and $b_{E}$ performs a measurement in the computational basis.
The third button $b_{R_x}$ performs $R_x(\frac{\pi}{2})$, and the final button $b_{\delta t}$ waits for a discrete time $\delta t$ (free evolution).
By applying $b_{\delta t}$ a total of $n$ times, we can wait time $t = n \cdot \delta t$.

Next, we choose a prior distribution from which to sample to begin Bayesian inference.
For convenience, these priors are summarized in \autoref{tab:ramsey_oqt}. 
As explained in \autoref{sec:setting_priors}, our prior is defined initially in a gauge-dependent way, which is then used to induce a prior distribution on the gauge-independent operational representation.
The initial $R_x\left(\frac{\pi}{2}\right)$ are sampled from a distribution that encompasses over- and under-rotation: we choose rotations of the form $R_x\left(\frac{\pi}{2} \pm \delta \theta\right)$, where $\delta \theta$ is a deviation sampled from a normal distribution with mean 0 and a small variance $\sigma^2 = 10^{-3}$. 
As $\delta t$ is meant to indicate evolution around the $z$ axis, we sample from $R_z(\theta)$ with $\theta$ chosen uniformly from between 0 and 1. 

For both the state preparation and measurement priors, we apply depolarization to the ideal state $\ket{0}$. 
When acting on a density matrix $\rho$, depolarization of strength $p$, $0 \leq p \leq 1$, sends 
\begin{equation}
 \rho \rightarrow (1 - p) \rho + \frac{p}{3} \left( X \rho X + Y \rho Y + Z \rho Z \right).
\end{equation}
The associated Bloch vector then transforms according to \cite{nielsen2002}:
\begin{equation}
 (a_x, a_y, a_z) \rightarrow ((1-p)a_x, (1-p)a_y, (1-p)a_z).
\end{equation}
Here all super-operators are expressed in the Pauli basis, where applying depolarization to super-operator $G$ sends
\begin{equation}
 G \rightarrow \begin{pmatrix}
                1 & 0 & 0 & 0 \\ 0 & 1-p & 0 & 0 \\ 0 & 0 & 1-p & 0 \\ 0 & 0 & 0 & 1-p
               \end{pmatrix} G.
\end{equation}
For the priors, we assume that depolarization occurs with strength $p$ chosen from the uniformly at random between 0 and 0.1, denoted $\mathcal{U}(0, 0.1)$.

The last step is to choose a set of fiducial sequences. 
We choose $\vec{f} = \{(\cdot), \quad (b_{R_x}), \quad (b_{R_x}, b_{R_x}), \quad (b_{R_x}, b_{\delta t}, b_{R_x})\}$.
These sequences are read from left to right; the first is the empty sequence, and application of SPAM is implicit in all sequences. 
This choice is not unique, and we picked some that performed well in practice.
Using these fiducials to construct an operational representation results in 27 parameters, which is reduced due to duplication from the 52 that are expected from counting the full $\tilde{E}, \tilde{F}$, and $\tilde{G}^k$. 

We initialized a SMC cloud with 10000 particles, and performed Bayesian inference over these parameters by feeding in simulated experimental data for sequences of the form $(b_{R_x}, \enskip (b_{\delta t})^n, \enskip b_{R_x})$ for $n = 2,\ldots 49$. 
The `true' values of the parameters that generated the data were randomly sampled from the prior distribution, with specific parameters given in \autoref{tab:ramsey_oqt}.
See the supplementary materials for the implementation.

In \autoref{fig:ramsey_likelihood} we plot the likelihood as calculated over the posterior distribution, and compare to the true likelihood (in this case, the set of model parameters that was used to produce the experimental data). 
We see that our inference has learned this operational representation, and produces comparable likelihoods even out to sequences that are double the length of those we trained with. Using \autoref{fig:ramsey_likelihood}, it is possible to fit a curve of the form $\cos^2(\omega t/2)$ and extract an estimate for the value of $\omega$. We obtain $\hat{\omega} = 0.345905$, a roughly $0.6\%$ difference from the true value $\omega = 0.346754$ as noted in \autoref{tab:ramsey_oqt}.

\begin{figure}[ht]
    \centering
        \includegraphics[width=0.48\textwidth]{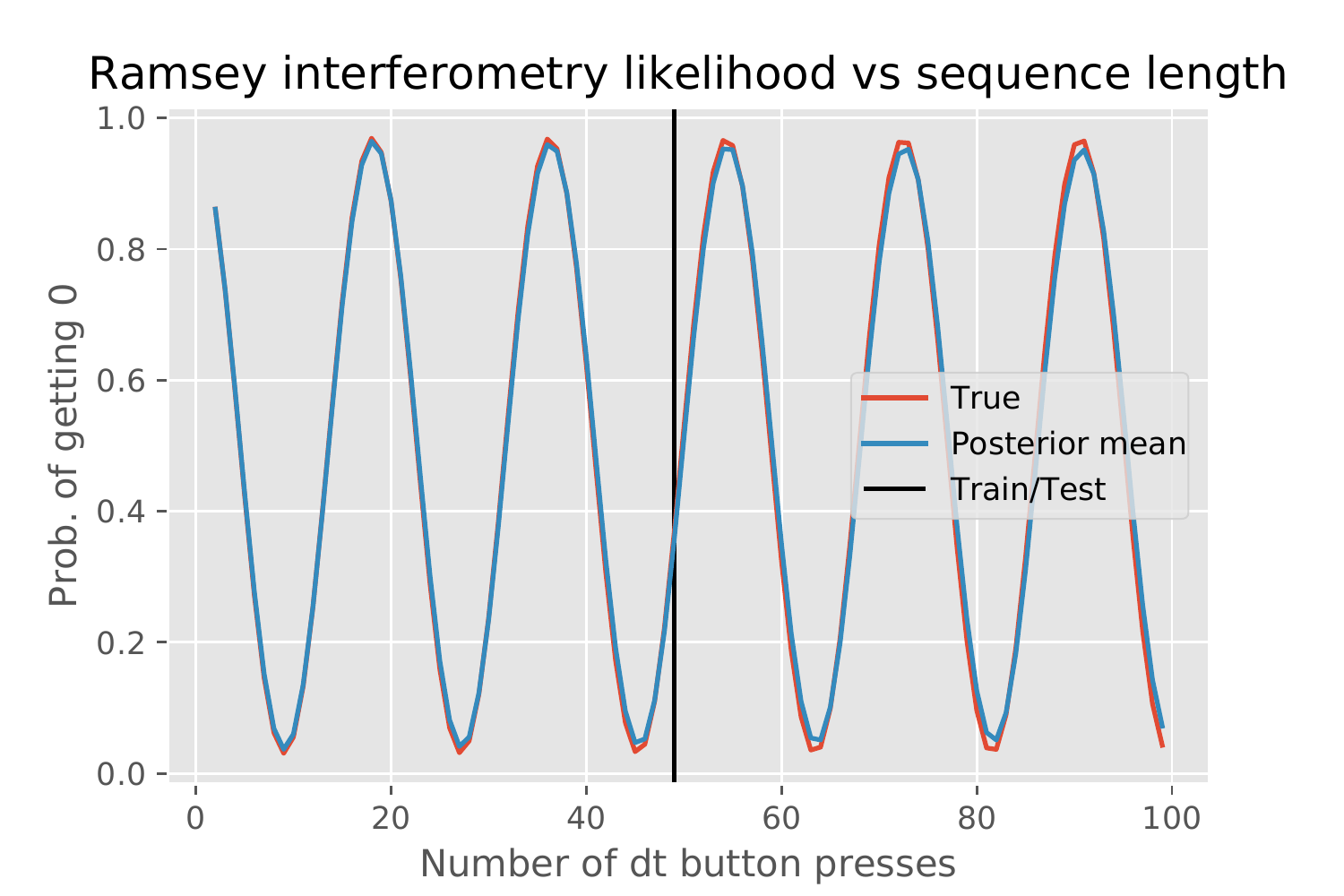}
        \includegraphics[width=0.48\textwidth]{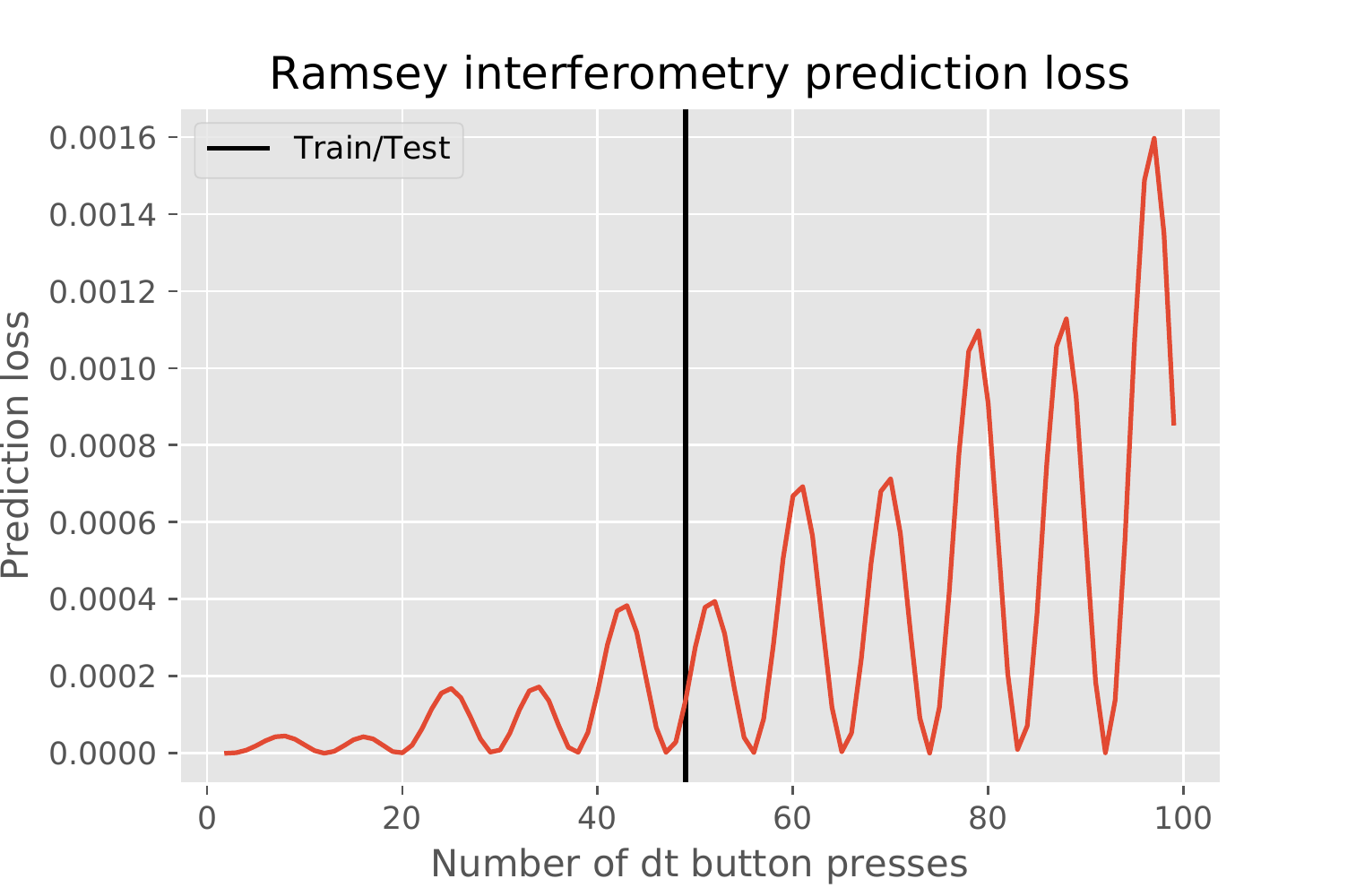}

    \caption{(Top) Likelihood vs sequence length for the true gate set compared to the gate set obtained by taking the mean over the posterior distribution. The mean posterior matches closely up to the testing point, and then begins to diverge. Fitting the curve produced $\hat{\omega} = 0.345905$, which is a roughly $0.6\%$ difference from the true value. (Bottom) The divergence can be quantified using a prediction loss. Shown here is the quadratic loss, $(\hat{p}_s - p_s)^2$. While small, it increases steadily as the sequence length increases.}\label{fig:ramsey_likelihood}
\end{figure}

Instead, we can judge the quality of the reconstruction by plotting prediction loss, as shown in the right panel of \autoref{fig:ramsey_likelihood}. 
The amount of quadratic loss is small in the absolute sense, and clearly worsens with experiment length, with the peaks increasing quadratically. 
To build upon \autoref{sec:predloss}, we include in the supplementary materials a similar plot using the KL divergence.

\begin{figure*}[tb]
    \centering
        \includegraphics[width=0.9\textwidth]{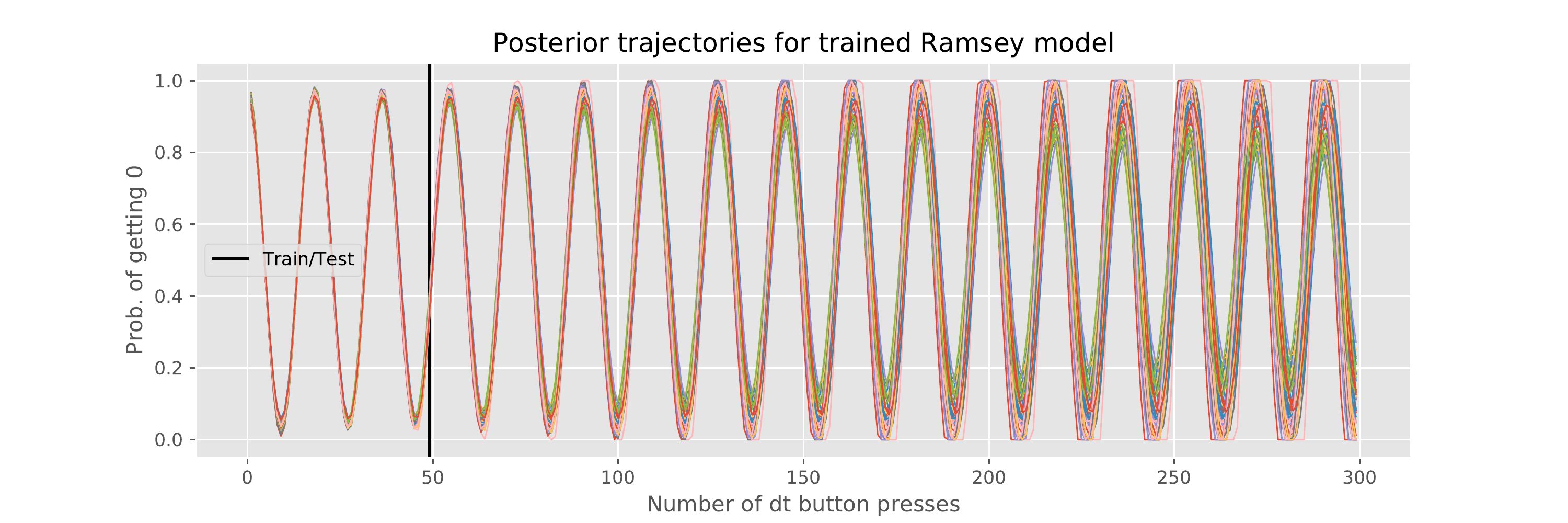}
        \caption{The trajectories of 50 particles sampled from the posterior operational representation for Ramsey interferometry. We see an increased spread at higher sequence lengths, which is highlighted later in \autoref{fig:ramsey_floating}.}
    \label{fig:ramsey_traj}
\end{figure*}

\begin{figure*}[tb]
    \centering
        \includegraphics[width=0.9\textwidth]{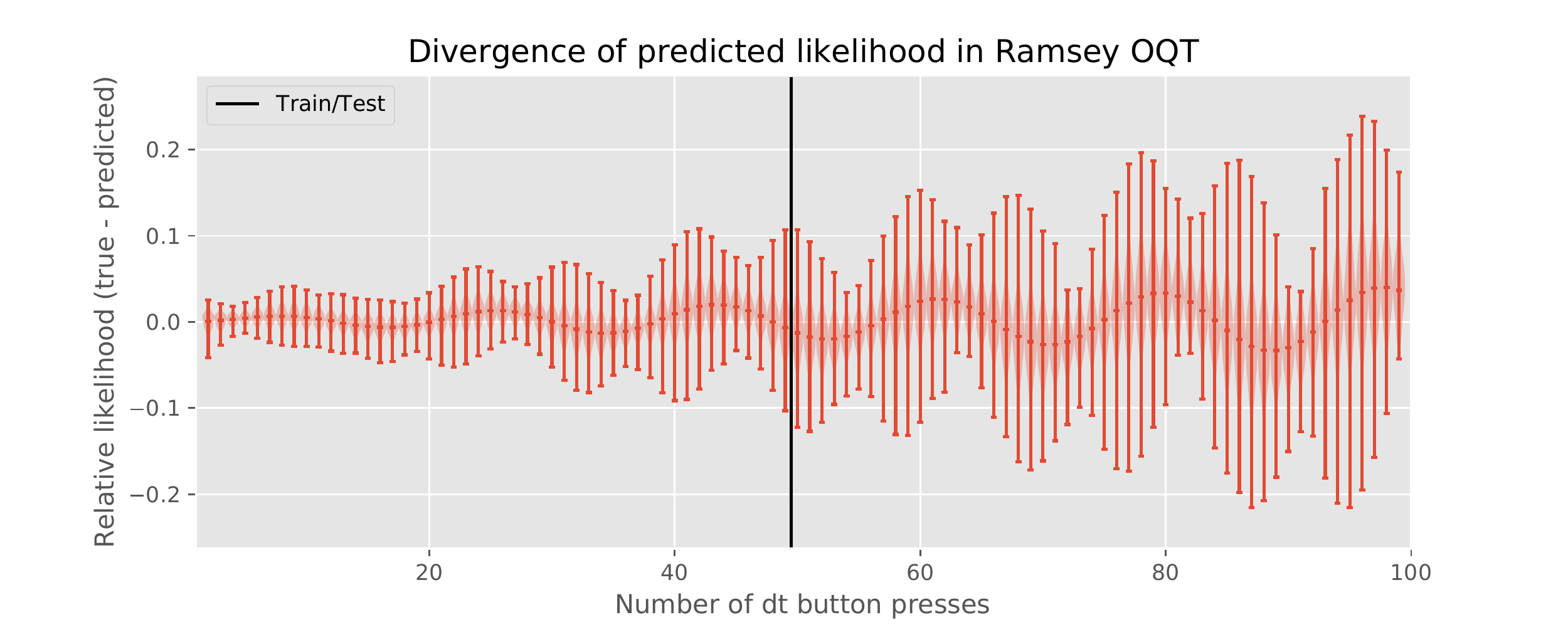}
        \caption{As the sequence length increases, the spread of likelihoods increases as well. Shown in this violin plot is the distribution of 'relative likelihood' for particles in the posterior distribution at each sequence length, i.e. $(\hat{p}_s - p_s)$.
    }
    \label{fig:ramsey_floating}
\end{figure*}

We can also visually examine the loss by plotting trajectories of different particles sampled from the posterior. 
Shown in \autoref{fig:ramsey_traj} are the trajectories of 50 such particles, with likelihoods computed out to sequences of up to $n=300$ presses of $b_{\delta t}$. 
As one might expect, we can see that the trajectories begin to `spread' significantly past the $n = 100$ point. 
The spread can also be quantified and visualized in the manner of \autoref{fig:ramsey_floating}, in which we have plotted the difference between the likelihoods of all particles of the posterior and the true likelihoods at each sequence length. 
We can see that the mean deviation from the likelihood increases as the spread in possible values becomes greater at longer sequence length.

While Ramsey interferometry is an arguably simple characterization procedure, it is perhaps the most surprising successful application of OQT we have explored.
The same task would not be possible in the traditional GST formalism if one is limited to performing only Ramsey-type experiments, namely two $R_x(\frac{\pi}{2})$ pulses separated by some amount of time. 
Circuits of that form are not rich enough to generate all the sequences required by GST. 
While one \emph{can} construct an informationally complete fiducial set using only compositions of
$R_x(\frac{\pi}{2})$ and $R_z(\delta t)$ gates, there will always be GST-required circuits that do not follow the Ramsey circuit form.  
(For example, GST will require at least one circuit that requires \emph{three} applications of $R_x(\frac{\pi}{2})$; such a circuit is not allowed if one is \emph{only} performing Ramsey circuits, all of which have only two applications of $R_x(\frac{\pi}{2})$.)  
Even though such circuits appear in the operational representation, our prior information allows us to not perform them if we so choose; this highlights the value of being able to incorporate prior information into a characterization protocol.  
Since the entire prior distribution is created computationally, we can still perform OQT even in cases where we are not able to physically perform a set of experiments that corresponds to every sequence in the operational representation.
\subsection{Long-sequence gate set tomography}

We next compare OQT to long-sequence GST, where carefully designed sequences are used to self-consistently fit both SPAM and an unknown gate set \cite{BlumeKohoutDemonstrationqubitoperations2017}.
Long-sequence GST uses the linear-inversion step of LGST as a starting point, and then proceeds with a longer maximum-likelihood estimation over experiments with progressively longer sequences of gates.
Once the procedure finishes, a final gauge fixing is often used to compare the resulting super-operators to expected super-operators. 

In  \cite{BlumeKohoutDemonstrationqubitoperations2017}, long-sequence GST was performed on experimental data from a trapped-ion qubit on which we could perform three operations: $G_i$, $G_x = R_x\left(\frac{\pi}{2}\right)$, and $G_y = R_y\left(\frac{\pi}{2}\right)$. 
Thus including SPAM, our box has five buttons.

The linear inversion step in  \cite{BlumeKohoutDemonstrationqubitoperations2017} was originally performed using 6 fiducials. 
However, choosing the same 6 fiducials here results in a $6 \times 6$ $\tilde{F}$ that has rank 4. 
The reason to include those extra fiducials is to increase stability, since LGST then represents an overdetermined system of linear equations.
In OQT, we can still include these extra experiments in our analysis, but since the fiducials are used directly to define our model parameters, we need to pick a subset of of size 4 (we choose $\vec{f} = \{(\cdot), \enskip (b_{G_x}), \enskip (b_{G_y}), \enskip (b_{G_x}, b_{G_x})\}$).

We perform OQT using the set of experiments included in the supplementary material of  \cite{BlumeKohoutDemonstrationqubitoperations2017}.
These experiments have the form $(\vec{f}_i, (\vec{g}_k)^L, \vec{f}_j)$, where $\vec{g}_k$ are `germ' sequences that are  specified in \autoref{tab:lsgst_parameters}. 
The particular germs were chosen in \cite{BlumeKohoutDemonstrationqubitoperations2017} because they are \emph{amplificationally complete}. From these experiments, we do not include those of the form  $((b_{G_x})^n)$, $((b_{G_y})^n)$, and $((b_{G_i})^n)$ for $n = 1, 2, 4, \ldots 8192$ in our training data set -- these are kept as a testing set.

The choice of prior plays a particularly important role here, due to the inherently noisy nature of a physical system.
We choose a very general prior, based on convex combinations of the ideal super-operators with ones chosen uniformly at random. 
For both $\rho$ and $E$, we take a combination of the form
\begin{equation}
 \rho^\prime = (1 - \epsilon)\rho + \epsilon \sigma, \quad \epsilon = 10^{-4}, \quad \sigma \in \hbox{GinibDM(2)}.
\end{equation}

GinibDM(2) denotes the Ginibre distribution, the uniform distribution over single-qubit density matrices.
Such states are sampled by computing 
\begin{equation}
 \sigma = \frac{X X^\dag}{\Tr(X X^\dag)}, \quad X_{ij} = a + bi, \enskip a, b \in \mathcal{N}(0, 1), 
 \label{eq:ginibre_qubit}
\end{equation}
where here $X$ is a $2 \times 2$ matrix. 
We take a similar approach for $G_i, G_x,$ and $G_y$ by adding Ginibre noise to the ideal super-operators:
\begin{equation}
 G' = (1 - \epsilon) G + \epsilon \Lambda, \quad \epsilon = 10^{-4}, \quad \Lambda \in \hbox{BCSZ}(2).
\end{equation}
Here, $\Lambda$ is a super-operator chosen from the uniform distribution over CPTP super-operators, known as the BCSZ distribution \cite{Bruzda2009}, denoted by BCSZ(2).

\begin{table*}
    \caption{
        \label{tab:lsgst_parameters}
        OQT parameters for long-sequence GST on trapped-ion data. Button labels are abbreviated $b_{G_x} \rightarrow G_x$ for simplicity here when specifying button sequences. All priors involve adding random noise to the original super-operators using the Ginibre distribution over qubits for $\rho$ and $E$ (denoted here by $\hbox{GinibDM}(2)$), and the Ginibre distribution for the super-operators (denoted by $\hbox{BCSZ}(2)$). The set of training experiments is the same as in \citet{BlumeKohoutDemonstrationqubitoperations2017}, however as denoted below we have removed a subset of these for testing.
    }
     \centering
        \begin{tabular}{p{0.2\textwidth}p{0.4\textwidth}p{0.3\textwidth}}
        \hline \hline
            \textbf{Button label} & \textbf{Prior} & \textbf{Example values} \\
            $\rho$ & $(1 - \epsilon) \ket{0}\bra{0} + \epsilon \sigma, \quad \sigma \in \hbox{GinibDM}(2)$ & $\epsilon = 10^{-4}$\\
            $G_x$ & $(1 - \epsilon) R_x(\frac{\pi}{2}) + \epsilon \Lambda, \quad \Lambda \in \hbox{BCSZ}(2) $ & $\epsilon = 10^{-4}$ \\
            $G_y$ & $(1 - \epsilon) R_y(\frac{\pi}{2}) + \epsilon \Lambda, \quad \Lambda \in \hbox{BCSZ}(2) $ & $\epsilon = 10^{-4}$ \\            
            $G_i$ & $( 1 - \epsilon) \openone +  \epsilon \Lambda, \quad \Lambda \in \hbox{BCSZ}(2)$ &  $\epsilon = 10^{-4}$ \\
            $E$ & $(1 - \epsilon) \ket{0}\bra{0} + \epsilon \sigma, \quad \sigma \in \hbox{GinibDM}(2)$ & $\epsilon = 10^{-4}$ \\
            \hline \hline 
            
    \textbf{Fiducial seqs.} & $\{(\cdot), \enskip (G_x), \enskip (G_y), \enskip (G_x, G_x)\}$ & \\
    \textbf{Training exps.} & $(\vec{f}_i,  (\vec{g}_k)^{L_k}, \vec{f}_j)$ for all fiducials and 11 germs $\vec{g}_k \in$ $\{(G_x)$, $(G_y)$, $(G_i,G_x, G_y)$,  $(G_x, G_y, G_i)$, $ (G_x, G_i, G_y)$, $(G_x, G_i, G_i)$, $(G_y, G_i, G_i),$  $ (G_x, G_x,  G_i, G_y)$,  $(G_x, G_y, G_y, G_i),$ $(G_x, G_x, $ $G_y, G_x, G_y, G_y)\}$ (unique sequences only, with testing sequences removed) & $L_k = \left\{ \left\lfloor \frac{2^m}{|g_k|}  \right\rfloor \right\}, m = 1, ..., 13$    \\
    \textbf{Testing exps.} & $((G_x)^n)$, $((G_y)^n)$, and $((G_i)^n)$ & $n = 1, 2, 4, ..., 8192$  \\
    \hline \hline
      \end{tabular}
\end{table*}

The choice of $\epsilon$ was informed by a combination of the experimental data and a grid search. 
Observing Figure 1c in \cite{BlumeKohoutDemonstrationqubitoperations2017}, we note that likelihoods in the (ideally) definite-outcome testing experiments start to significantly decay at around $10^4$ gates, hence we intuit that $\epsilon$ should be around $10^{-4}$. 
This was later confirmed using a grid search. 
We ran OQT using a cloud of 10,000 particles for 192 different combinations of $\epsilon$'s. 
We set $\epsilon$ the same for $G_i, G_x, G_y$ at $\{10^{-m}, \enskip 2\cdot 10^{-m}, \enskip 4\cdot 10^{-m}, \enskip 8\cdot 10^{-m} \}$ for $m = 5, 4, 3$. 
For SPAM, we also choose $\epsilon$ the same for $\rho$ and $E$, and explore the range $\{10^{-m}, \enskip 2\cdot 10^{-m}, \enskip 4\cdot 10^{-m},\enskip 8\cdot 10^{-m}\}$ for $m = 5, 4, 3, 2$. 

The quality of each pairing of $\epsilon$ was determined by (a) whether or not the SMC updater succeeded without all particle weights going to zero, and (b) the sum of the total variation distance over the testing experiments.
For a given sequence $s$, let $p(s, \mathcal{R})$ and $p(s, \mathcal{E})$ be the experimental probabilities for a given reconstruction $\mathcal{R}$ and the experimental data $\mathcal{E}$. 
The total variation distance (TVD) between the two probability distributions is:
\begin{equation}
 \hbox{TVD}(s, \mathcal{R}, \mathcal{E}) = |p(s, \mathcal{R}) - p(s, \mathcal{E})|.
\end{equation}

Bayesian inference ran to completion\footnote{We note that the larger values of $\epsilon$ for which inference did not complete in this case may still yield results if the number of particles is increased, given that the noisy super-operators obtained with smaller $\epsilon$ will still be in the support of the prior. This highlights the trade-offs one can explore between time, computational resources, and the strength of our assumptions about the buttons.} for $\epsilon$ of the gates in the range $4 \cdot 10^{-5}$ up to $2 \cdot 10^{-4}$. For these values, inference was successful over essentially the  full range of SPAM values. However the sum of total variation distances was notably lower for gate $\epsilon = 10^{-4}$, and SPAM $\epsilon$ between $10^{-5}$ and $10^{-4}$, reaching a minimum of $10^{-4}$ during one full sweep of the grid search. 

\begin{figure*}[ht]
    \centering
    \includegraphics[width=\textwidth]{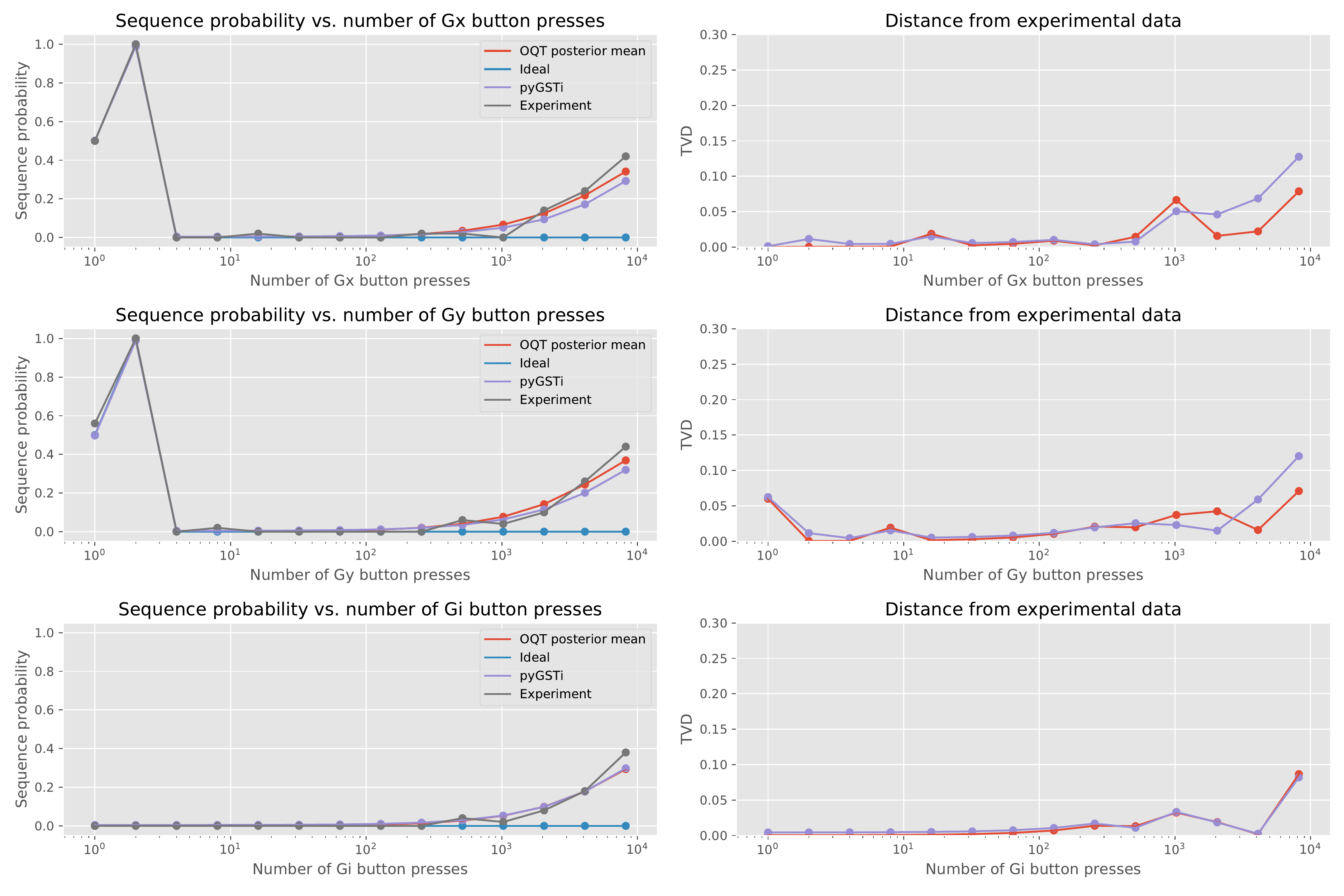}
    \caption{(Left) Comparison of OQT posterior likelihoods obtained using parameters in \autoref{tab:lsgst_parameters} to  ideal likelihoods, pyGSTi reconstruction, and
    experimental data for the trapped-ion data 
    of \cite{BlumeKohoutDemonstrationqubitoperations2017}. The testing experiments 
    consist of exponentially longer sequences of repeated button presses, $G_x^k$ 
    and $G_i^k$. (Right) Total variation distance of pyGSTi and OQT reconstructions for the same gate sequences. Here we see that the OQT results vary from those of pyGSTi for $G_x$ and $G_y$, but give comparable results for $G_i$. The total TVD for OQT across all testing experiments is lower, at 0.724, while that of pyGSTi is 0.961.}
    \label{fig:lsgst_likelihoods_and_tvd}
\end{figure*}

Results for OQT run with the parameters of \autoref{tab:lsgst_parameters} are plotted in \autoref{fig:lsgst_likelihoods_and_tvd}. 
The left column of plots compares the likelihoods predicted for the test sequences from the OQT posterior distribution to the likelihoods of the 'perfect' gate, the experimental counts, and the gate set reconstructed in \cite{BlumeKohoutDemonstrationqubitoperations2017} using the pyGSTi software package. 
We see that OQT produces results that are competitive with its contemporaries without the need to perform MLE.
This is quantified in the right column that plots the variation distance for the same set of experiments.
The total TVD for the OQT posterior is 0.724, and that of the pyGSTi reconstruction is 0.961.

\subsection{Randomized benchmarking}

Like GST, OQT equips us to make predictions about the outcome of any future experimental sequences.
Hence, as has been done before using GST \cite{BlumeKohoutDemonstrationqubitoperations2017}, we can use  OQT to perform randomized benchmarking (RB).
To do so, we perform OQT to learn the generators of the Clifford group.
Then, using samples taken from the obtained posterior, we will apply RB type sequences and compute the survival probability. 

\subsubsection{Background for randomized benchmarking}

RB makes use of random elements of the Clifford group, which for one qubit is constructed using two generators, $\mathcal{C} = \langle H, S \rangle$, where
\begin{align}
    H = \frac{1}{\sqrt{2}} \begin{pmatrix} 1 & 1 \\ 1 & -1 \end{pmatrix}, 
    \quad\text{and}\quad
    S = \begin{pmatrix} 1 & 0 \\ 0 & i \end{pmatrix}.
\end{align}
Up to a phase, $\mathcal{C}$ contains 24 elements.

A traditional RB experiment seeks to characterize the errors present in our Clifford gates. 
We begin by preparing a known state $\rho_\psi$, and then apply a randomly chosen sequence of Clifford elements.
This is followed by applying the element that is the inverse of the group element
formed by the sequence (not just performing the sequence backwards). We then measure our system using the measurement operator $E_\psi$ 
corresponding to our initial state.

If there are no errors in the Clifford gates, the action of the sequence and its inverse would cancel, leaving the
state exactly as we found it. When there are errors, however, we can compute what is termed the \emph{survival probability}
of the original state. As the sequences increase in length, the survival probability decays, as errors accumulate.
Typically, one plots a ``decay curve'' of the form
\begin{align}
    \label{eq:rb-mean}
    P(m) = (A - B) p^m + B,
\end{align}
where $m$ is the sequence length, and where $P(m)$ is the mean survival probability over all sequences of length $m$.
That is, we define that
\begin{align}
    P(m) \defeq \expect_{\vec{s} \in  \text{s. t.} |\vec{s}| = m} \left[\Pr(0 | [\Phi]; \vec{s}) \right].
\end{align}
We note that since probabilities are \emph{not} directly observable, and can only be estimated, caution must be taken when estimating $P(m)$ or interpreting estimates obtained in an \emph{ad hoc} fashion.

Keeping this caution in mind, the form \autoref{eq:rb-mean} for the expectation value of the survival probability over sequences of length $m$ was derived analytically by \citet{MagesanCharacterizingQuantumGates2012}, where it was noted that the parameter $p$ contains information about the average fidelity of our Clifford elements.
In particular,
\begin{equation}
    p = \frac{d F_{\ave}(\Lambda) - 1}{d - 1},
\end{equation}
where $d$ is the dimension of the Hilbert space under consideration ($d = 2$ for a single qubit RB experiment), where $F_{\ave}(\Lambda)$ is the \emph{average gate fidelity} of the channel $\Lambda$, and where $\Lambda$ is the average error in implementing each member of the Clifford group.
In particular, $\Lambda$ takes on the gauge-dependent form
\begin{align}
    \Lambda = \expect_{U \sim \mathcal{C}} [(U^\dagger \bullet {}) \Lambda_U],
\end{align}
where $(U^\dagger \bullet {})$ is the ideal action of $U^\dagger$, and $\Lambda_U$ is a super-operator representing the actual implementation of $U$.

Despite the large literature on RB \cite{emerson2005scalable, emerson2007symmetrized, knill2008randomized, magesan2011scalable, MagesanCharacterizingQuantumGates2012, carignan2015characterizing, cross2016scalable, brown2018randomized, hashagen2018real, helsen2017multiqubit, mckay2017three, epstein2014investigating, ProctorWhatrandomizedbenchmarking2017, WallmanRandomizedbenchmarkinggatedependent2017, proctor2019direct}, 
both the experimental implementation and statistical interpretation are challenging.
Since RB is frequently used to assess suitability for quantum error correction applications, this is troubling.
Since the technique that we describe here \emph{indirectly} performs RB through \emph{ex post facto} simulation, we are less vulnerable to some of these challenges.
More details on this can be found in Appendix~\ref{apx:rb}.
\subsubsection{Performing RB using OQT}

\begin{table*}
    \caption{
        \label{tab:rb_parameters}
        OQT parameters for randomized benchmarking. Button labels are abbreviated  $b_H \rightarrow H$ and $b_S \rightarrow S$ for simplicity when specifying button sequences.
    }
        \begin{tabular}{p{0.146\textwidth}p{0.6\textwidth}p{0.18\textwidth}}
        \hline \hline
            \textbf{Button label} & \textbf{Prior} & \textbf{Example values} \\
            $\rho_\psi$ & $1/\sqrt{2} \begin{pmatrix} 1 & 0 & 0 & 1 \end{pmatrix}$ & Perfect \\
            $H$ & $(1-\epsilon) G_H(\delta \theta_H) + \epsilon \Lambda_H$, $\enskip \delta \theta_H \in \mathcal{N}(0, 0.0015), \enskip \Lambda_H \in \hbox{BCSZ}(2)$ & Eqs. \autoref{eq:hadamard_prior},\autoref{eq:rb_h_sample}; $\newline \epsilon = 10^{-3}$ \\
             $S$ & $(1-\epsilon) R_z(\pi/2 + \delta\theta_S) + \epsilon \Lambda_S, \enskip \delta \theta_S \in \mathcal{N}(0, 0.0015), \enskip \Lambda_S \in \hbox{BCSZ}(2) $ & Eq. \autoref{eq:rb_s_sample}; $\newline \epsilon = 10^{-3}$ \\
           
            $E_\psi$ & $1/\sqrt{2} \begin{pmatrix} 1 & 0 & 0 & 1 \end{pmatrix}$ & Perfect \\ \hline \hline
    \textbf{Fiducial seqs.} & $\{(\cdot),\enskip (H),\enskip (H, S, H),\enskip (S, H, S)\}$ & \\
    \textbf{Training exps.} & 100 random RB sequences of increasing length $n$ & $n = 40, \ldots 60$ \\
    \textbf{Testing exps.} & 100 random RB sequences at each of 87 logarithmically spaced $n$ & $n = 10, \ldots, 252$  \\
    \hline \hline
        \end{tabular}
\end{table*}

To perform RB using OQT, we need a box with 4 buttons: $\rho_\psi, E_\psi, b_H,$ and $b_S$. 
As RB is robust to SPAM errors, we assume for simplicity that SPAM is perfect.
That is, we focus on learning $H$ and $S$. 
Our first step is to choose an appropriate prior over $H$ and $S$: we pick one that represents our belief that errors in each generator are due to both systematic over-rotations and Ginibre noise.
To apply over-rotation to the Hadamard, we begin with its super-operator representation $G_H  = H \otimes H$.
Mathematically, this can also be written as $H \otimes H = e^{\frac{i\pi}{2} \left(H \otimes \id - \id \otimes H \right)}$, which we recognize as just evolution for the time $\pi/2$ under the Lindbladian $L = (\id \otimes H - H^{\T} \otimes \id)$.
We perturb the evolution time slightly to write
\begin{multline}
 \label{eq:hadamard_prior}
 G_H(\delta \theta) 
 = e^{i \left( \frac{\pi}{2} + \delta \theta \right) \left( H \otimes \id - \id \otimes H \right) } \\
 = \cos^2 (\delta \theta) H\otimes H + \sin^2 (\delta \theta) \id \otimes \id + \\ \frac{i}{2} \sin(2 \delta \theta) \left( \id \otimes H - H \otimes \id  \right). 
\end{multline}
An $S$ gate is simply a $R_z \left(\frac{\pi}{2} \right)$ gate, so in line with the previous examples, we  choose a distribution $R_z \left(\frac{\pi}{2} + \delta \theta \right)$ where $\delta \theta$ is normally distributed with mean 0 and variance $\sigma^2_\theta$.

We then add Ginibre noise to both $H$ and $S$ by sampling random super-operators from the BCSZ distribution, such that the sampled super-operators will have the form
\begin{eqnarray}
 G_{H} &\rightarrow& (1 - \epsilon) G_H\left(\delta \theta_H\right) + \epsilon \Lambda_H \quad \text{and} \\
 G_{S} &\rightarrow& (1 - \epsilon) R_z\left(\frac{\pi}{2} + \delta \theta_S\right) + \epsilon \Lambda_S.
\end{eqnarray}

For the presented example, we chose $\epsilon = 10^{-3}$, and $\delta \theta_H, \delta \theta_S \in \mathcal{N}(0, \sigma^2 = 0.0015)$. 
With respect to the choice of $\sigma^2$, it can be shown that a channel over-rotated by $\delta \theta$ has fidelity $F = \frac{2}{3} + \frac{1}{3} \cos(2 \delta \theta)$.
Fidelities on the order of 0.999 are typical of qubits today, and so assuming that $\delta \theta$ corresponds roughly to the standard deviation, we choose $\sigma^2 = 0.0015$.
As we are assuming the addition of Ginibre noise, the actual fidelity of our operations will be slightly lower than this.

To generate data, we chose a `true' version of $G_H$ and $G_S$ by sampling from
the prior. The sampled parameters, as listed in the supplementary materials, are
\begin{eqnarray}
 \delta \theta_H &=& -0.007798, \quad
 \delta \theta_S = -0.047391,\\ \nonumber
  \Lambda_H &=&
 \setlength\arraycolsep{1.7pt}
  \begin{pmatrix}                                   
 1     &  0    &   0  &     0     \\
 0.435103 &-0.120449 &-0.297836 & 0.062722 \\
-0.314789 & 0.032982 &-0.089239 & 0.080124 \\
 0.188424 & 0.101214 &-0.284711 & 0.142465
 \end{pmatrix} \label{eq:rb_h_sample} \\ \nonumber
 \Lambda_S &=& 
 \setlength\arraycolsep{1.7pt}
 \begin{pmatrix}
 1     &   0   &    0 &        0 \\   
 -0.256911 & 0.53382 &  0.265858 & 0.104777 \\
 -0.018402 & -0.178172 & 0.565879 & 0.061297 \\
 -0.187707 & -0.349921& -0.279835 & 0.450564
 \end{pmatrix} \label{eq:rb_s_sample}
\end{eqnarray}
where the super-operators $\Lambda_H, \Lambda_S$ are expressed in the Pauli basis.

With this prior distribution, we initialized a cloud of 10000 particles. 
Bayesian inference was performed to learn $G_H$ and $G_S$ by training with 100 RB sequences of length 40 to 60, using an equal number of sequences at each length.
We then tested the model using 87 sequence lengths logarithmically spaced from the range from 10 to 252, using 100 random sequences at each length.
For each particle in the posterior distribution, we compute the survival probability  for each sequence (the same set of sequences was used for each particle). 
For each particle we can then fit to a curve of the form $P(m) = (A - B) p^m + B$ to obtain the traditional set of RB fit parameters. 
The parameters $A, B$ are constrained to be between 0 and 1, and $p$ to be between $-0.5$ and 1. 
The fit is a least-squares fit weighted by variance, since at every sequence length the survival probability is averaged over 100 different sequences.

\begin{figure}[tb]
    \centering
    \includegraphics[width=\columnwidth]{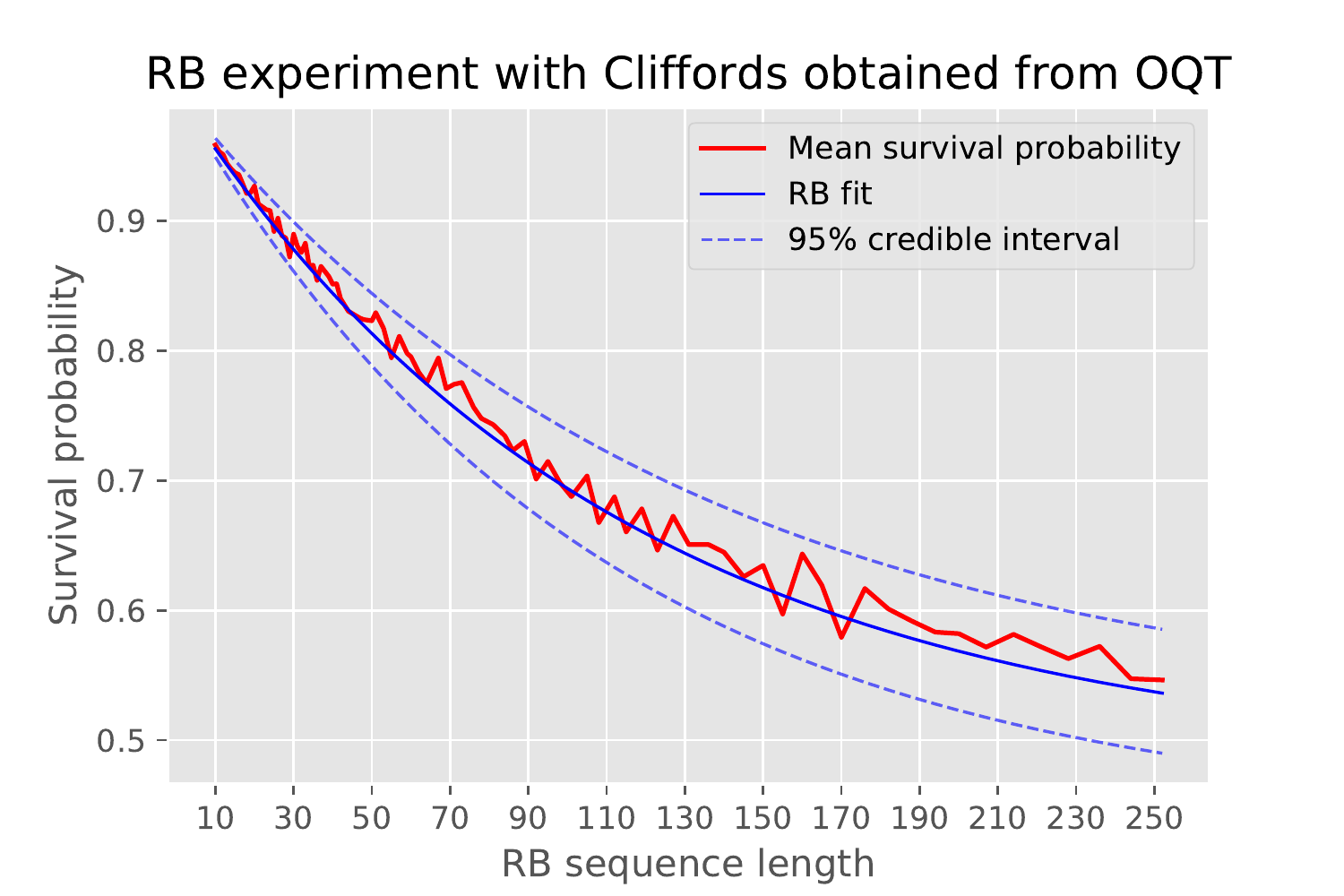}
    \caption{RB decay curve for our learned Clifford group, with 95\% credible interval.  The survival probability of the $y$-axis represents the average over RB sequences of a fixed length; the mean survival probability on the plot is the survival probability at length $m$ averaged over 100 experiments, on which we then take the weighted average over the full posterior distribution. We fit a curve for each particle, and display here the curve that is the weighted average of the fit parameters. The fit has the form $P(m) = (A-B)p^m + B$, with mean parameters $(A, B, p) = (0.999916, 0.481494, 0.991119)$, and thus average gate set fidelity 0.995560. The `true' average gate set fidelity falls within the computed credible interval of $[0.995304, 0.996115]$.}
    \label{fig:RB_confidence}
\end{figure}

The mean survival probability is shown in \autoref{fig:RB_confidence}.
At each length, the mean is computed first for each particle over the set of 100 sequences, and then a weighted average over these means is taken using the particle weights in the posterior distribution.
Since each particle yields a set of $(A, B, p)$, we can also compute the weighted mean of these parameters, shown as the solid blue line in \autoref{fig:RB_confidence}.
The mean fit parameters are $(A, B, p) = (0.999916, 0.481494, 0.991119)$
The mean value of $p$ can be used to compute an average gate set fidelity of $(1 + p)/2 = 0.995560$.
Using the same testing experiments, we can compute the RB decay rate for the `true' gate set that generated the data.
We obtain a `true' value of $0.995337$. 

We can plot the 95\% credible interval over all RB parameters using the Bonferroni correction \cite{bonferroni1936, dunn1961bonferroni}.
This interval is shown in \autoref{fig:RB_confidence} as dotted lines.
We obtain for $p$ the interval $[0.990608,  0.992230]$, corresponding to fidelities of $[0.995304, 0.996115]$, which neatly contains the `true' value of $0.995337$.
Details and additional plots are available in the supplementary materials.

\section{Quantum mechanics in an operational representation}
The previous examples showed a wide range of different characterization tasks that can be implemented within an operational representation.  However, a further foundational question is what the limits of this formalism are, and whether or not all of quantum mechanics can be understood within our formalism.  One of the fundamental challenges about the gateset tomography model that we inherit is that the description of the gateset is entirely discrete.  That is to say that the device in question contains a set of buttons.  However, quantum dynamics is naturally continuous.  Therefore, understanding quantum dynamics within the language of GST is challenging.  Here, we will show that we can think of such continuous dynamics as yielding an equation of motion for the gateset of a quantum system.  This not only shows that operational GST is general enough to describe all of quantum mechanics, but also provides a new way of modelling quantum dynamics using only gauge-independent parameters (i.e. observable quantities) and thereby eschewing the use of unobservable quantities such as quantum state operators which appear in conventional treatments of quantum dynamics.


In particular, to allow arbitrary quantum dynamics it is convenient to think now of our operational representation as being an explicit function of time.  We assume here for simplicity that the gates, fiducials and measurements all are given by time-independent sequences.  As an example, let us consider the case where $\partial_t\sket{\rho(t)} = \mathcal{L} \sket{\rho(t)}$, where $\mathcal{L}$ is the Lindbladian super-operator and $\sket{\rho(t)}$ is the initial state evaluated for the system at time $t$.

The equation of motion for the operational representation of the gate set is then
    \begin{eqnarray}
        \partial_t \tilde{E}_i(t) &=& \sbra{E} F_i \partial_t\sket{\rho(t)} = \sbra{E} F_i \mathcal{L}\sket{\rho(t)}, \nonumber\\
        \partial_t \tilde{F}_{ij}(t) &=&  \sbra{E} F_i F_j \partial_t\sket{\rho(t)}= \sbra{E} F_i F_j \mathcal{L} \sket{\rho(t)}, \\
        \partial_t \tilde{G}^{(k)}_{ij}(t) &=& \sbra{E} F_i G_k F_j \partial_t \sket{\rho(t)}=\sbra{E} F_i G_k F_j \mathcal{L} \sket{\rho(t)},  \nonumber
    \end{eqnarray}
A challenge with this representation is that its evaluation relies on objects that we do not know \emph{a priori} and are not related (directly) to observed quantities since the expectation values of $\mathcal{L}$ are not assumed to be known in the operational representation.  In part, this has to do with the way that we have chosen to represent $\mathcal{L}$.  In the following, let us assume that there exist coefficients $\alpha_\ell$ such that
\begin{equation}
\mathcal{L} = \sum_\ell \alpha_\ell F_\ell.\label{eq:lindbladL}
\end{equation}
These values of $\alpha_\ell$ can further be learned empirically using the operational representation.  Let us assume that we empirically measure by choosing $\delta\ll 1$ and taking $\partial_t\widetilde{E}(t)\approx (\widetilde{E}(t+\delta) - \widetilde{E}(t))/\delta$.  If we then take the resultant vector to be $\dot{\widetilde{{\bf{E}}}}(t)$, $\widetilde{\bf{F}}(t)$ to be the matrix representation of the $\widetilde{F}_{ij}(t)$ tensor and take $\boldsymbol{\alpha}$ to be the unknown matrix of coefficients for the Lindbladian using~\eqref{eq:lindbladL} then if $\widetilde{\bf F}$ is an invertible matrix then
\begin{equation}
\dot{\widetilde{{\bf{E}}}}(t) = \widetilde{\bf{F}}(t) \boldsymbol{\alpha} \Rightarrow \boldsymbol{\alpha} = \widetilde{\bf{F}}^{-1}(t)\dot{\widetilde{{\bf{E}}}}(t).
\end{equation}
Thus such a representation can be learned if $\widetilde{\bf{F}}(t)$ is an invertible matrix.  If not, a least-squares approximation can be found by applying the Moore-Penrose pseudoinverse in its place.  Of course, this merely proves the existence of a solution (or a least-squares solution) for the coefficients of the Lindbladian as a function of the Fiducials.  In practice, Bayesian methods such as the ones considered here and elsewhere may be of great use for both learning and quantifying the uncertainty in the model Lindbladian.

Given a set of coefficients for the Lindbladian the first order system of equations that governs the evolution can be expressed as
    \begin{eqnarray}
        \partial_t \tilde{E}_i(t) &=&\sum_\ell \alpha_\ell\sbra{E} F_i F_\ell \sket{\rho(t)}, \nonumber\\
        \partial_t \tilde{F}_{ij}(t) &=&  \sum_\ell \alpha_\ell \sbra{E} F_i F_j F_\ell \sket{\rho(t)}, \\
        \partial_t \tilde{G}^{(k)}_{ij}(t) &=& \sum_\ell \alpha_\ell \sbra{E} F_i G_k F_j F_\ell \sket{\rho(t)}, 
    \end{eqnarray}
As we can see the derivatives of $\widetilde{E}_i(t)$ depend on the values of $\widetilde{F}_{ij}(t)$ but the derivatives of $\widetilde{F}_{ij}$ and $\widetilde{G}_{ij}^{(k)}$ depend on expectation values of cubic functions of the fiducials.  Thus we can solve these equations, but doing so may require more information in some cases.  Below we consider two important cases.  The first is where the set of fiducial super-operators is not closed under multiplication and the second is where the group is closed and and consists of at most quadratic polynomials in the fiducials.

    \subsection{Dynamics for infinite sets of fiducials}
As a first example of how the dynamics of the operational representation works, consider the case where the fiducial super-operators form an infinite group wherein the group product is given by super-operator multiplication.  In this case, we cannot assume any structure to the fiducials that will cause products of them to contract to a finite set of super-operators.

If we have such a model then the dynamics can again be written in terms of a set of observables, however the set that needs to be measured becomes larger in this setting.  In particular, we extend the definition of the $\widetilde{E}$ and $\widetilde{G}$ tensors such that
\begin{align}
\tilde{E}_{i_1,\ldots,i_p}(t) &= \sbra{E} F_{i_1} \cdots F_{i_p} \sket{\rho(t)}.\nonumber\\
\tilde{G}_{ij_1,\ldots,j_p}^{(k)}(t) &=  \sbra{E} F_i G_k F_{j_1} \cdots F_{j_p} \sket{\rho(t)}.
\end{align}
Under these assumptions the dynamics of the operational representation of the gate set takes the form of a driven first order dynamical system.
    \begin{eqnarray}
        \partial_t \tilde{E}_i(t) &=&  \sum_{\ell} \alpha_\ell \tilde{F}_{i\ell}(t), \nonumber\\
        \partial_t \tilde{F}_{ij}(t) &=& \sum_{\ell} \alpha_\ell \tilde{E}_{ij\ell}(t), \nonumber\\
&\vdots&\nonumber\\
		\partial_t \tilde{F}_{i_{1}\ldots i_{n}}(t) &=& \sum_{\ell} \alpha_\ell \tilde{E}_{i_{i}\ldots i_{n}\ell}(t), \nonumber\\
&\vdots&\nonumber\\
        \partial_t \tilde{G}^{(k)}_{ij}(t) &=& \sum_{\ell} \alpha_\ell \tilde{G}_{ij\ell}^{(k)}(t)\nonumber\\
&\vdots&\nonumber\\
        \partial_t \tilde{G}^{(k)}_{ij_{1}\ldots j_{n}}(t) &=& \sum_{\ell} \alpha_\ell \tilde{G}_{ij_{1}\ldots j_{n}\ell}^{(k)}(t)\nonumber\\
&\vdots
    \end{eqnarray}
Thus the entire dynamics of the gate set can be predicted if the $\tilde{E}$ and $\tilde{G}$ tensors are known in their entirety.  This is operationally equivalent to the Schr\"{o}dinger equation, while eschewing the need for unobservable quantities such as the quantum state.  While solving the resultant dynamical equations formally requires knowing an infinite hierarchy of terms to predict future dynamics perfectly, in many cases the super-operators for the fiducials will form a finite group making knowledge of the complete hierarchy of tensors unnecessary.

Finally, in practice the entire hierarchy is not needed in order to accurately estimate the dynamics for all subsequent times from data at a single time given the decomposition of the Lindbladian into a sum of fiducials. We have from Taylor's theorem and Stirling's approximation that
\begin{multline}
\left| \tilde{E}_{i}(t+\Delta)  - \sum_{j=0}^K \frac{\partial_t^j \tilde{E}_i(t) \Delta^j}{j!}\right| \le \frac{(\sum_{\ell} |\alpha_\ell|\Delta)^{K+1} }{(K+1)!} \\
\le \left(\frac{(\sum_{\ell} |\alpha_\ell|\Delta)}{K+1}\right)^{K+1}.
\end{multline}
Thus by solving this equation for the value of $K$ that yields error $\epsilon$ we find that a sufficient value of $K$ is
\begin{equation}
K= \left\lfloor\frac{\ln(1/\epsilon)}{\text{LambertW}\left(\frac{\ln(1/\epsilon)}{(\sum_{\ell} |\alpha_\ell|\Delta)}\right)}\right\rfloor \in O\left( \frac{\ln(1/\epsilon)}{\ln(\ln(1/\epsilon)) }\right),
\end{equation}
if $\Delta \le \sum_\ell |\alpha_\ell|$.  Thus the total number of terms needed to simulate the dynamics for a short time step with error at most $\epsilon$. varies logarithmically with the error tolerance.  Each such term can be approximated using Monte-Carlo sampling such that the estimate of the derivatives is at most $\epsilon$ using a number of samples that scales as $O({\rm poly}(1/\epsilon))$ and therefore even in the case where the algebra does not close the dynamics can be simulated using a small number of observables.  It should be noted that in the event that the fiducials form a closed group that this scaling improves exponentially Monte-Carlo sampling is no longer required to estimate the derivatives.  This shows that under reasonable assumptions the operational representation can also be used to describe the dynamics of a quantum system that we can probe experimentally using a set of fiducial operations and gates.  Hence, while inspired by problems of characterization in quantum systems, much broader classes of quantum dynamical problems can also be discussed using our formalism while only making reference to observable quantities.

\subsection{Dynamics for closed sets of fiducials}
Next let us consider a simpler case where the set of fiducial super-operators is closed under multiplication.  Specifically, let $S = \{F_i \bigcup F_iF_j\}$ be the set of all monomials and binomials in the fiducials.  Next because the set is closed under multiplication there exists a function $g$ such that for any $s_i$ and $s_j$ in $S$ there exists $s_{g(i,j)}$ such that $s_i s_j = s_{gf(i,j)}$.  Also for simplicity, assume that the sets are laid out in lexicographic ordering such that $s_1 =F_1, s_2 = F_2,\ldots$.  It then follows that if we use the fact that the set is closed then the equations of motion for the operational representation greatly simplify to the following finite system of equations
\begin{eqnarray}
        \partial_t \tilde{E}_i(t) &=&\sum_\ell \alpha_\ell\sbra{E} F_i F_\ell \sket{\rho(t)} = \sum_{\ell} \alpha_\ell \widetilde{F}_{i\ell}(t),\nonumber\\
        \partial_t \tilde{F}_{ij}(t) &=&  \sum_\ell \alpha_\ell \sbra{E} s_{g(i,g(j,\ell))} \sket{\rho(t)}, \nonumber\\
        \partial_t \tilde{G}^{(k)}_{ij}(t) &=& \sum_\ell \alpha_\ell \sbra{E} F_i G_k F_j F_\ell \sket{\rho(t)}=\sum_\ell \widetilde{G}_{ij\ell}^{(k)}(t) , \nonumber\\
\partial_t \tilde{G}^{(k)}_{ij_1j_2}(t) &=& \sum_\ell \alpha_\ell \sbra{E} F_i G_k s_{g(j_1,g(j_2,\ell))} \sket{\rho(t)}\label{eq:closedDynamics}
    \end{eqnarray}
These equations can, in many cases be solved directly without having to truncate (as was done in the infinite case considered above).  Also, because of the lack of a curse of dimensionality the resulting equations can be solved within error $\epsilon$ using $O({\rm polylog}(1/\epsilon)$ operations via existing differential equations solvers.  For this reason, cases where the fiducial operators form a closed (or approximately closed) set under multiplication can greatly simplify the equations of motion.  However, it should be noted that such cases are highly restrictive and, for example, preclude the inclusion of depolarizing noise or similar effects because such noise models will typically lead to a fiducial set that is not closed under multiplication.  For such situations, truncating the infinite dynamics at finite order may be preferable.

\subsection{Generator Learning}
A further observation is that because the dynamics considered above can be represented as a set of coupled first-order differential equations, we can use this representation to think of generalize the notion of Hamiltonian learning beyond the framework originally proposed.  In particular, consider the dynamics in~\eqref{eq:closedDynamics}.  Let us consider a concatenation of all such terms in~\eqref{eq:closedDynamics} of the form

\begin{equation}
\Psi(t) = [\vec{\tilde{E}}(t), \vec{\tilde{F}}(t), \vec{\tilde{G}}_{ij}^{(k)}(t),\vec{\tilde{G}}_{ij\ell}^{(k)}(t)]^T,
\end{equation} 
where we explicitly include the indices of $\vec{G}$ above to differentiate the rank 2 and rank 3 $\vec{G}$ tensors.  We then have from the theory of differential equations and the fact that~\eqref{eq:closedDynamics} is a homogeneous first-order differential equation that there exists a generator $\vec{K}$ such that for any condition $\Psi(0)$,

\begin{equation}
\partial_t \Psi(t) = K \Psi(t),~\qquad \Psi(t) = e^{Kt} \Psi(0).
\end{equation}
This means that we can also infer dynamical models for a gateset using Bayesian inference.  In particular, if we have an initial description of our gateset $\Psi(0)$ then evolve some time $t$ and after applying a gate sequence then we would have that the posterior distribution can be expressed as
\begin{equation}
P(K|E) = \frac{P(E| K; \Psi(0)G_0,\ldots,G_{N-1},t) P(K)}{\int P(E| K; \Psi(0)G_0,\ldots,G_{N-1},t) P(K)\mathrm{d}K}
\end{equation}
This model ends up assuming that the time required for the gate sequence to be implemented is negligible compared to the dynamical timescale for the gateset.  In the event that the timescales are comparable, then $K$ only becomes the instantaneous generator of time-displacements for the gateset and the result will become an ordered operator exponential rather than the simple operator exponential given above.

The key point behind these observations is that techniques that are more reminiscent of quantum Hamiltonian learning (such as in~\cite{Granade_rohl_2012}) can also be included within our operational representation.  This not only shows that the framework is broader than it may have first appeared but also that we can apply the same ideas employed in that literature in order to infer models for the dynamics of a gateset.  This allows some forms of non-Markovian noise to be incorporated in our models without leaving the operational representation.

\section{Conclusions}

We have demonstrated a framework for quantum tomography in which we can represent many other characterization tasks. 
Working with a gauge-independent representation of the system, we can learn its behavior from experimental data and predict the outcomes of future experiments. 
OQT gives us the freedom to incorporate prior information computationally (without any physical experiments). 
Future improvements to OQT involve the extension to two-qubit operations, as well as allowing for buttons to be held down for arbitrary duration (i.e. time-dependent operations).

\begin{acknowledgments}
CG thanks Chris Ferrie for teaching him the contents of Appendix~\ref{apx:review-bayes-est} over the years. ODM is grateful to the quantum group at Microsoft for numerous visits and the opportunity to participate in an internship. TRIUMF receives funding  via a contribution through the National Research Council of Canada. IQC is supported in part by the Government of Canada and the Province of Ontario.

Sandia National Laboratories is a multimission laboratory
managed and operated by National Technology \& Engineering Solutions of Sandia, LLC, a wholly owned subsidiary of
Honeywell International Inc., for the U.S. Department of Energy’s National Nuclear Security Administration under contract DE-NA0003525. This research was funded, in part, by
the U.S. Department of Energy, Office of Science, Office of
Advanced Scientific Computing Research Quantum Testbed
Program. This paper describes objective technical results and
analysis. Any subjective views or opinions that might be expressed in the paper do not necessarily represent the views
of the U.S. Department of Energy or the United States Government.

This research was supported by PNNL’s Quantum Algorithms, Software, and Architectures (QUASAR) LDRD Initiative. 
\end{acknowledgments}

\nocite{apsrev41Control}
\bibliographystyle{apsrev4-1}
\bibliography{apsrev-control,bayesian-gst}

\begin{appendices}

\section{Review of Bayesian estimators}
\label{apx:review-bayes-est}

In this Appendix, we provide a brief review of estimation theory as applied to Bayesian inference.
In doing so, it is convenient to first consider estimation more generally.
Suppose that there is some vector $\vec{x} \in \mathcal{X}$ of parameters that we would like to learn given some data $D \in \mathcal{D}$, where $\mathcal{X}$ is the set of feasible values for $\vec{x}$, and where $\mathcal{D}$ is the set of data we could have possibly obtained.
Then, we will say that any function $\hat{\vec{x}}(\cdot) : \mathcal{D} \to \mathcal{X}$ which accepts data and returns estimates is an \emph{estimator}.

For example, given any $\vec{x}_0 \in \mathcal{X}$, the constant function $\hat{\vec{x}}(D) = \vec{x}_0$ is an estimator that disregards all evidence in favor of returning $\vec{x}_0$.
Clearly, while this is a valid estimator, it is not a very \emph{good} one to use in practice.
Our task in estimation theory is then to recommend a particular estimator that is desirable according to some set of practical considerations.
We may, for example, want an estimator that incurs as little error as possible.

We can formalize this desire by introducing a function $L : (\mathcal{X} \times \mathcal{X}) \to \mathbb{R}^+$ such that $L(\hat{\vec{x}}, \vec{x})$ is the \emph{loss} that we incur if we return $\hat{\vec{x}}$ as our estimate when the true value is $\vec{x}$.
For example, if we are estimating a single real number ($\mathcal{X} = \mathbb{R}$), then we may choose the squared error $L(\hat{x}, x) = (\hat{x} - x)^2$ as our loss.
More generally, for $\mathcal{X} = \mathbb{R}^d$ for $d \in \mathbb{N}$, the quadratic loss $L_{\matr{Q}}(\hat{\vec{x}}, \vec{x}) = (\hat{\vec{x}} - \vec{x})^\T \matr{Q} (\hat{\vec{x}} - \vec{x})$ is a well-defined loss function for any positive definite matrix $\matr{Q}$.

Once we have decided upon a loss function, we can then reason about what losses we may incur in a given experiment using a particular estimator.
To do so, we first need to extend our definition of loss from estimates to estimators by taking the average over all possible data sets that an estimator could take as input.
Concretely, given a loss function $L$, define the \emph{risk} $R : (\mathcal{D} \to \mathcal{X}) \to \mathbb{R}^+$ of an estimator as
\begin{align}
    R(\hat{\vec{x}}, \vec{x}) \defeq \expect_{D \sim \Pr(D | \vec{x})}[L(\hat{\vec{x}}(D), \vec{x})].
\end{align}

The risk implicitly defines a multi-objective optimization problem, in that an estimator that works well for a particular ground truth need not work well more generally.
At an extreme, the constant estimator $\hat{\vec{x}}(D) = \vec{x}_0$ works beautifully well when $\vec{x} = \vec{x}_0$.
We thus at a minimum want an estimator that minimizes the risk that we incur in some case of interest.
To formalize this notion, we say that an estimator $\hat{\vec{x}}(\cdot)$ is \emph{dominated} by an estimator $\hat{\vec{x}}'(\cdot)$, if for all $\vec{x}$, $R(\hat{\vec{x}}, \vec{x}) \ge R(\hat{\vec{x}}', \vec{x})$, and if there exists some $\vec{x}$ for which this inequality is strict.
Put differently, an estimator dominates another estimator if it is less risky in all circumstances, such that there is no decision-theoretic basis for preferring the dominated estimator.
An estimator which is not dominated by any other estimator is said to be \emph{admissible}.

From a Bayesian perspective, however, we are generally most interested in minimizing what we expect the risk to be given our experience with a system so far.
We can make this precise by taking the expectation value of the risk with respect to a prior distribution to obtain the \emph{Bayes risk} of an estimator,
\begin{align}
    r(\hat{\vec{x}}) \defeq \expect_{\vec{x} \sim \Pr(\vec{x})}[R(\hat{\vec{x}}, \vec{x})].
\end{align}
The unique estimator minimizing the Bayes risk for a particular loss function is called the \emph{Bayes estimator} for that loss,
\begin{align}
    \hat{\vec{x}}_{\Bayes} \defeq \operatorname{arg\,min}_{\hat{\vec{x}}(\cdot)} r(\hat{\vec{x}}(\cdot)).
\end{align}
By construction, the Bayes estimator is admissible: any estimator that dominates the Bayes estimator would have a strictly smaller Bayes risk.
Under fairly weak conditions \cite{kiefer1953}, however, we can conclude the converse as well, namely that every admissible estimator is the Bayes estimator for a particular prior distribution.

In full generality, computing the Bayes estimator for a particular loss function requires minimizing over functions of all data sets, which is not feasible or practical.
Some loss functions, however, allow for much more efficiently computing Bayes estimators.
In particular, \emph{Bregman divergences} are loss functions which can be written as the difference between a convex function and its first-order Taylor expansion.
If a loss function is Bregman, then the celebrated theorem of \cite{bgw_optimality_2005} shows that
\begin{align}
    \hat{\vec{x}}_{\Bayes}(D) = \expect_{\vec{x}}[\vec{x} | D].
\end{align}
That is, the posterior mean of $\vec{x}$ is the Bayes estimator for any Bregman divergence.

Many practically relevant loss functions are Bregman divergences, including the squared error, quadratic loss, and Kullback--Liebler divergence.
Thus, the posterior mean gives us a method of efficiently computing admissible estimators that minimize the average error we incur in inference procedures.
As we saw in \autoref{subsec:particle-filter}, the posterior mean can be efficiently computed using particle filtering, giving us a practical method for reporting Bayes estimates.

\section{Quantum state tomography}
\label{app:statetomo}

\begin{table*}
    \caption{
        \label{tab:statetomo_oqt}
        OQT parameter specification for rebit state tomography. Button labels are abbreviated $b_{R_x} \rightarrow R_x$ for notational simplicity. State tomography was performed independently for 1000 states (and associated gate sets) sampled from the distributions below.  As such, we do not provide examples of the sampled parameters in this case. For all priors, the value of $p$ is the amount of depolarization.
    }

        \begin{tabular}{p{0.146\textwidth}p{0.45\textwidth}p{0.32\textwidth}}
            \hline \hline
            \textbf{Button label} & \textbf{Prior} & \textbf{Example values} \\
            $\rho$ & Ginibre rebit distribution, Eq.\autoref{eq:ginibre_rebit}, \enskip $p \in \mathcal{U}(0, 0.1)$ & 1000 randomly selected states\\
            \texttt{$R_x$} & $R_x(\pi/2 + \epsilon), \enskip \epsilon \in \mathcal{N}(0, 10^{-3})$, \enskip $p \in \mathcal{U}(0, 0.1)$ &  \\
            \texttt{$R_y$} & $R_y(\pi/2 + \epsilon), \enskip \epsilon \in \mathcal{N}(0, 10^{-3}) $, \enskip $p \in \mathcal{U}(0, 0.1)$ &  \\
            $E$ & $1/\sqrt{2} \begin{pmatrix} 1 & 0 & 0 & 1 \end{pmatrix}$, \enskip  $p \in \mathcal{U}(0, 0.1)$ &  \\ \hline \hline
    \textbf{Fiducial seqs.} & $\{(\cdot), \enskip (R_x), \enskip (R_y), \enskip (R_x, R_x)\}$ & \\
    \textbf{Training exps.} & 50 randomly chosen products of $n$ fiducials & $n = 1, \ldots, 10$; $n$ increases linearly\\
    \textbf{Testing exps.} & 50 randomly chosen products of $n$ fiducials &  $n = 5, \ldots, 15$; $n$ increases linearly
        \\ \hline \hline
        \end{tabular}
\end{table*}

\begin{figure*}[tb]
 \centering
 \includegraphics[width=\textwidth]{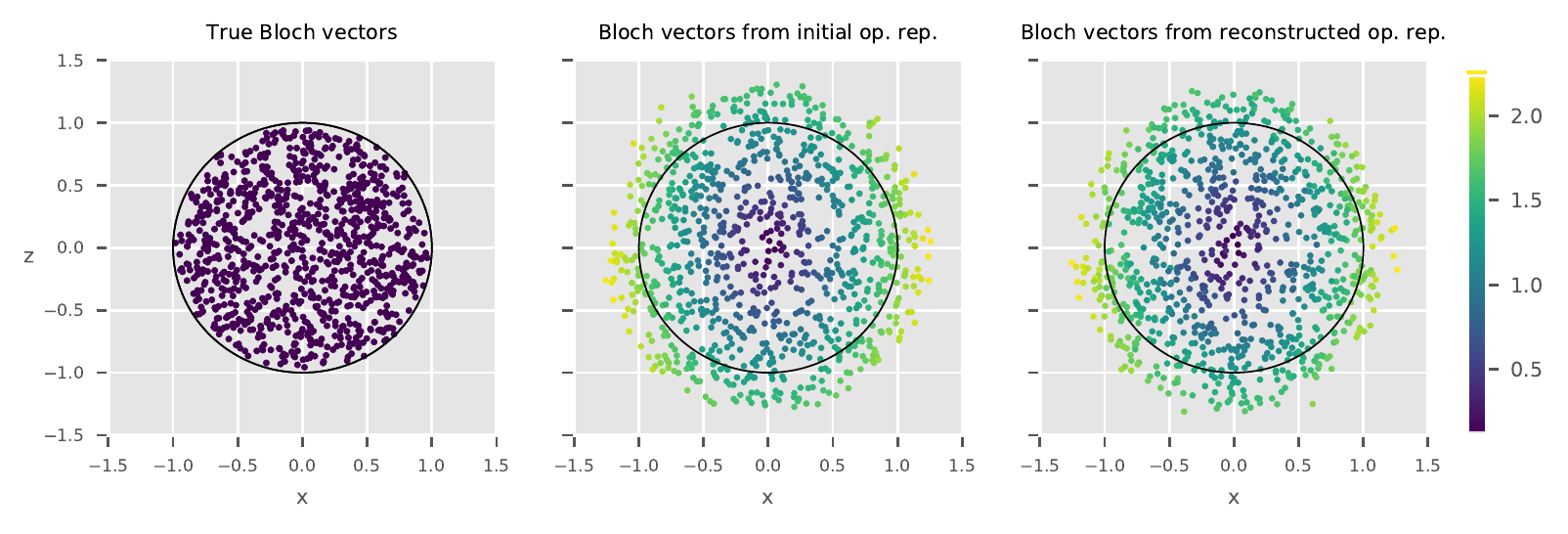}
 \caption{What happens when we perform state tomography `naively' using the measurement results from noisy gates we had assumed were perfect. (Left) The 1000 initial random states, sampled from the prior. They are rebits, and have only $x$ and $z$ components. (Center) A `pseudo Bloch circle' constructed by pulling coordinates from the initial operational representation, i.e. fiducial experiment probabilities, as per \autoref{eq:bloch_coordinates}. Points are colored by their distance to the corresponding true states in the left panel. (Right) The same plot as for the middle, but calculating the Bloch coordinates using the posterior mean after performing OQT. See \autoref{fig:naive_pseudovector_comparison} for a histogram of the colored difference before and after reconstruction.}
 \label{fig:pseudo_rebits}
\end{figure*}
In traditional quantum state tomography, we seek to learn an unknown state using a set of measurements that are presumed to be perfectly known. 
Naively performing this task, however, leads to estimates that are not self-consistent. 
We provide a demonstration of this by performing OQT on unknown rebits, i.e. qubits with no $y$-components.

As in previous examples, the first step is to phrase the problem in terms of our operational formalism.
We will consider the case where our box again has 4 buttons: two SPAM buttons, and ones that we believe perform  $R_x \left( \frac{\pi}{2} \right)$ (denoted $b_{R_x}$) and $R_y \left( \frac{\pi}{2} \right)$ (denoted $b_{R_y}$).
We add uncertainty to our rotation buttons by setting the priors for $R_x \left( \frac{\pi}{2} \right)$ and $R_y \left( \frac{\pi}{2} \right)$ to be over-rotations with a mean of 0 and a variance of $10^{-3}$. 
We also add depolarization to the rotation gates, with strength $p \in \mathcal{U}(0, 0.1)$. 
 
We sample our states from the Ginibre rebit distribution,  the uniform distribution over rebit states.
Such states are sampled by computing 
\begin{equation}
 \rho = \frac{X X^\dag}{\Tr(X X^\dag)}, \quad X_{ij} \in \mathcal{N}(0, 1), 
 \label{eq:ginibre_rebit}
\end{equation}
where in our case, $X$ is a $2 \times 3$ matrix\footnote{In the more general case, the Ginibre distribution of $d \times d$ density matrices is sampled by populating a $d \times d$ matrix $X$ with complex values $a+bi$ where both $a$ and $b$ are randomly sampled from $\mathcal{N}(0, 1).$ For random real density matrices, we must sample matrices of size $d \times (d+1)$ \cite{Osipov2010}.}. 
The rebit states are subject to a small amount of depolarization with strength $p \in \mathcal{U}(0, 0.1)$. 
We apply similar depolarization to the measurement $E = \ket{0} \bra{0}$. 
Full details of our parameter specifications are shown in \autoref{tab:statetomo_oqt}. 
 
The set of chosen fiducial sequences is $\vec{f} = \{(\cdot), \enskip (R_x), \enskip (R_y), \enskip (R_x, R_x)\}$.
If our buttons were perfect, this set of fiducials provides a set of measurements that is informationally complete in the traditional sense, meaning that the measurements span the entire Bloch sphere. 
However in practice these will be noisy - our definition of informationally complete thus shifts to whether or not the fiducials produce a well-conditioned $\tilde{F}$; we find that the chosen set is reliable in practice.

In the `naive' method of performing state tomography, the fiducial sequences and associated probabilities would be directly related to the coordinates on the Bloch sphere $(a_x, a_y, a_z)$:
\begin{eqnarray}
 a_x = 2 p_{x} - 1, &\enskip& p_{x} = \tilde{F}_{02} = \Tr\left[\sket{\rho} \sbra{E} R_y(\pi/2)\right] \nonumber \\
 a_y = 2 p_{y} - 1, &\enskip& p_{y} = \tilde{F}_{01} = \Tr\left[\sket{\rho} \sbra{E} R_x(\pi/2)\right]  \label{eq:bloch_coordinates} \\
 a_z = 2 p_{z} - 1, &\enskip& p_{z} = \tilde{F}_{00} = \Tr\left[\sket{\rho} \sbra{E}\right] \nonumber
\end{eqnarray}
In the remainder of this section, we will demonstrate the consequences of this naive method.

\begin{figure*}[tb]
 \includegraphics[width=\textwidth]{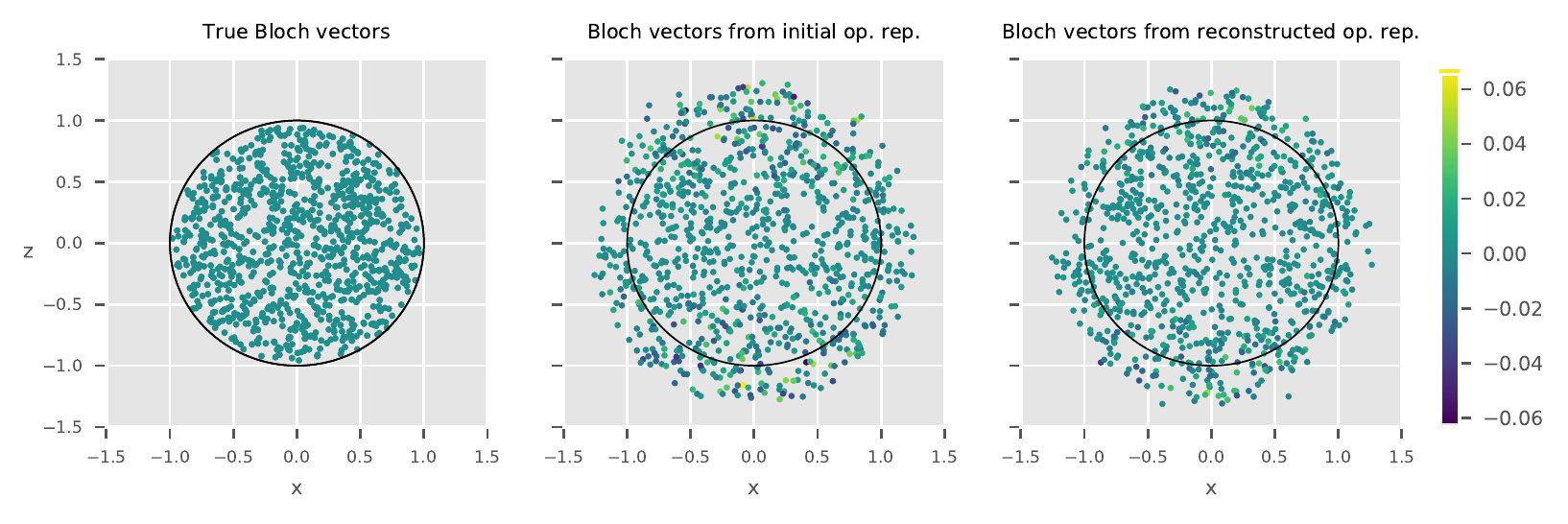}
 \caption{For the same set of states in \autoref{fig:pseudo_rebits}, we color the states according to the $y$ component of the pseudo Bloch vectors. In theory this should always be 0, but we observe here that our naive reconstruction method produces slight deviations both before and after reconstruction. However we note that after reconstruction, the deviation is less, as displayed in the right panel of \autoref{fig:naive_pseudovector_comparison}.}
 \label{fig:pseudo_rebits_ycolour}
\end{figure*}

\begin{figure*}[tb]
 \includegraphics[width=\textwidth]{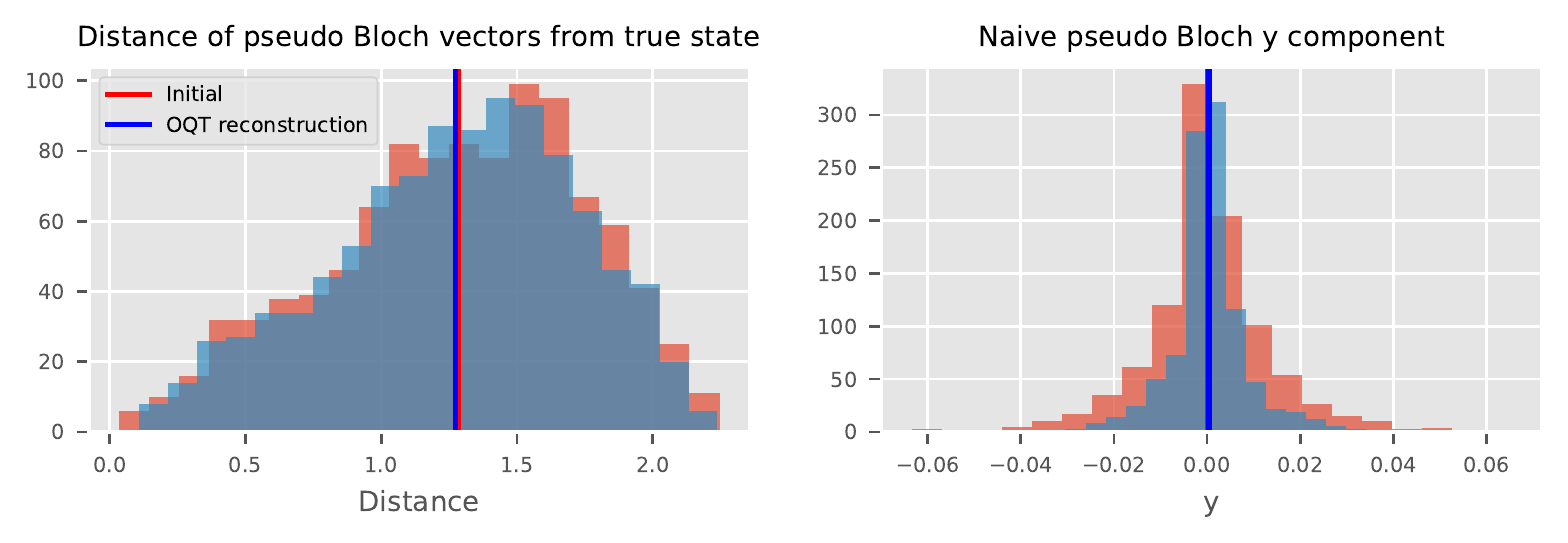}
 \caption{Histograms of pseudo Bloch vector properties before and after performing OQT. Solid lines show the mean of the corresponding distribution. OQT learns these vectors well and produces comparable distributions, but this naive method of tomography nevertheless leads to a noticeable $y$ component added to many of the rebits.}
 \label{fig:naive_pseudovector_comparison}
\end{figure*}

We performed state tomography with OQT independently on 1000 random states. 
In \autoref{fig:pseudo_rebits}, we have plotted the true Bloch coordinates of the initial states in the left panel.
In the middle panel, we see the coordinates obtained from their initial operational representations according to  \autoref{eq:bloch_coordinates}.
States pulled from the Ginibre ensemble should lie firmly within the boundaries of the Bloch sphere, or circle, in the rebit case.
However reconstruction according to \autoref{eq:bloch_coordinates} produces Bloch coordinates that fall well outside the boundaries.
Furthermore, they pick up small $y$-component, as demonstrated in the first two panels of \autoref{fig:pseudo_rebits_ycolour}. 

For our OQT experiments, we push the state preparation button once, then apply a sequence of randomly 
selected gate buttons from a minimum length of 1 to a maximum length of 10. 
We then measure, record the outcome, and repeat 50 cases to form a training corpus. 
The sequence length steadily increased during training, with the same amount of sequences generated at each length. 

We note that \autoref{fig:pseudo_rebits} illustrates the dangers of `naive' state tomography in the presence of measurement errors.
For each of the hypotheses shown in the middle and right-hand plots of \autoref{fig:pseudo_rebits}, if a naive tomographer were to take an infinite amount of data from a system described by that hypothesis and then reconstruct the initial state $\rho$, they would correctly conclude either that their data was impossible, or that $\rho$ lies outside the Bloch sphere entirely.
Put differently, if one assumes that the measurement sequences used in a state tomography experiment are ideal, then naive state tomography will return absurd results \emph{even} in the limit of infinite data.

\section{Details on randomized benchmarking}
\label{apx:rb}
In this appendix, we discuss the advantages of performing randomized benchmarking within an operational framework.

\subsection{Using operational formalisms to perform randomized benchmarking}
\citet{MagesanCharacterizingQuantumGates2012} derived that $A$ and $B$ contain information about the state preparation and measurement errors incurred by a randomized benchmarking experiment.
They are expressed analytically as
\begin{align}
    A & = \Tr \left(E_\psi \Lambda \left(\rho_\psi \right) \right), \\
    \text{and } B & = \Tr \left( E_\psi \Lambda \left( \frac{\openone}{d} \right) \right).
\end{align}
A key point here is that traditional RB assumes that $\Lambda$ is the 
same for all elements of the Clifford group.
However, as we will see,
Clifford elements implemented in the GST framework will naturally have different errors, as elements are
composed of sequences of $H$ and $S$ of varying lengths. 

If the implementations of each Clifford element are perfect, we obtain $A = 1$,  $B = 1/2$, and $p = 1$, and so the survival probability is identically 1 for all sequences. 
However in the worst case, we obtain something essentially depolarized and so $p = 0$, meaning that the curve will immediately decay to $B = 1/2$. Fitting the experimental data to a curve of this form can thus give an idea of the value of $p$, which in turn can give us an estimate of the average gate fidelity.  

Before proceeding, it is helpful to establish that, despite its apparent simplicity, learning figures of merit from randomized benchmarking data is an astonishingly subtle problem that warrants no small amount of caution.
Especially given the rigorous demands placed on randomized benchmarking results for application to predicting the success of fault-tolerance, it is of the utmost importance that the results of RB experiments are understood in full recognition of the caveats placed on said results by current experimental and theoretical limitations.
For instance, as mentioned above, for instance, the derivation of \citet{MagesanCharacterizingQuantumGates2012} rests \emph{critically} on the assumption that the noise on each element of a gate set is independent of which element is being considered.
While \citet{MagesanCharacterizingQuantumGates2012} does provide a derivation that attempts to include gate-dependence, later counterexamples have shown that this assumption cannot even be made in a gauge-independent fashion \cite{ProctorWhatrandomizedbenchmarking2017} --- this implies that the gate-independence assumption cannot be experimentally tested.
Later work has shown that the effects of gate-dependence exponentially small effects on randomized benchmarking data \cite{WallmanRandomizedbenchmarkinggatedependent2017, fong2017randomized}, but it is still an open question as to how to meaningfully interpret RB data.

Perhaps more pressing still, the original derivation of \citet{MagesanCharacterizingQuantumGates2012} only derived the \emph{mean} survival probability and not any higher moments.
A fitting procedure such as homoscedastic least-squares fitting (the default procedure offered by MATLAB, SciPy, and many other packages, see Appendix ~\ref{apx:rb-estimation} for a review) will thus necessarily give incorrect or misleading answers, as the \emph{variance} over randomized benchmarking data depends both on the variance within each sequence and over shots of that sequence, and on the variance between different sequences.
This challenge can be overcome by committing to taking exactly one repetition of each sequence before choosing a new sequence \cite{GranadeAcceleratedrandomizedbenchmarking2015}, but this is feasible only for a small number of experimental platforms, such as those controlled by custom FPGA firmware \cite{HeeresImplementingUniversalGate2016}.
As an alternative solution, one can introduce \emph{nuisance parameters} to track the unknown higher moments and estimate them at the same time as the expectation of interest.
A recent proposal of this form was advanced by \citet{HincksBayesianInferenceRandomized2018}, who introduced a parameterization for RB protocols that includes a distribution at each sequence length that is then sampled using Hamiltonian Monte Carlo, effectively introducing an uncountable number of nuisance parameters in a way that they can be efficiently estimated.

From this perspective, using OQT to analyze randomized benchmarking data provides an explicit and gauge-independent nuisance parameterization that avoids both the interpretational and practical difficulties of drawing inferences from RB data.
We can then rely on the procedure of \citet{BlumeKohoutDemonstrationqubitoperations2017}~to synthesize from a final posterior over operational representations RB data of a form that is immediately amenable to analysis by even relatively informal methods such as heteroscedastic least-squares fitting.

\subsection{Estimation within randomized benchmarking}
\label{apx:rb-estimation}

In this Appendix, we review the estimation theory underlying randomized benchmarking and summarize some of the most prevalent pitfalls.
To do so, we will rely heavily on the Likelihood Principle \cite{ste_bayesian_2007}, which informally states that in order to make decisions consistent with experimental observation, we must base our decisions only on the evaluation of a likelihood function at our data, and cannot base our inference on any property of our data that is not expressed in the likelihood.
For RB in particular, this consistency requirement forces us to describe our implementations of RB in an operational manner, such that we can write down likelihood functions.

For instance, we recall that as per \autoref{eq:rb-mean}, the \citet{MagesanCharacterizingQuantumGates2012} model gives us that the mean sequence probability
\begin{align}
    P(m) \defeq (A - B) p^m + B
\end{align}
for some parameters $\vec{y} = (p, A, B)$.
This is not yet an operational description, however, as sequence probabilities are \emph{not} observable properties of finite-length experiments \footnote{As an amusing aside, this realization implies that the word ``observable'' in many formulations of quantum mechanics is reserved for those objects which are fail to be observable. It is for this reason that we prefer the more operational description offered by the POVM formulation.}.
To make an operational description of the \citet{MagesanCharacterizingQuantumGates2012} model \autoref{eq:rb-mean}, let us be more precise about a description of our experimental procedure.
As a prototypical example of such a description, most RB experiments proceed as follows:

\begin{enumerate}
    \item Perform the following for each $m \in \{m_0, \dots, m_{M - 1}\}$:
    \begin{enumerate}
        \item Perform the following $N$ times:
        \begin{enumerate}
            \item Choose a random sequence $\vec{s}$.
            \item Perform the following $K$ times:
            \begin{enumerate}
                \item Prepare a state $\rho$.
                \item Apply the sequence $\vec{s}$
                \item Measure the POVM $\{E, \id - E\}$.
            \end{enumerate}
            \item Record the number of times that $E$ was observed in the above loop as $k(\vec{s})$.
        \end{enumerate}
        \item Record the mean of $k(\vec{s})$ for each $\vec{s}$ sampled in the above loop as $n(m_i)$.
    \end{enumerate}
\end{enumerate}
We recognize the innermost loop as being a sample from the binomial distribution
\begin{subequations}
    \label{eq:rb-inner-loop}
    \begin{align}
        k(\vec{s}) & {} \sim \Bin(\Pr(E | [\Phi]; \vec{s}), K), \\
        \Pr(k | \vec{s}) & {} = \sum_k {K \choose k} p_{\vec{s}}^k (1 - p_{\vec{s}})^{K - k},
    \end{align}
\end{subequations}
where we have taken the shorthand $p_{\vec{s}} \defeq \Pr(E | [\Phi]; \vec{s})$ to denote sequence probabilities of the form considered in the rest of the paper.
From the perspective of RB, however, this is problematic, as a sequence probability for the sequence $\vec{s}$ can in general depend on any element of the operational representation for $[\Phi]$.
We may not be able to compute the sequence probability $\Pr(E | [\Phi]; \vec{s})$ given only hypotheses about the RB parameters $\vec{y}$.

Nonetheless, the \citet{MagesanCharacterizingQuantumGates2012} model gives us hope that we may still be able to formulate a likelihood function for the entire experiment, even if we cannot do so for each individual sequence within an RB procedure.
Following this hope, let us marginalize \autoref{eq:rb-inner-loop} over the choice of sequence $\vec{s}$, since we have chosen $\vec{s}$ randomly at the start of our loop over sequences.
Concretely,
\begin{equation}
 \begin{split}  
    &\Pr(k | [\Phi]; |\vec{s}| = m_i) \\ 
    &\enskip =  \expect_{\vec{s} \text{ s.t. } |\vec{s}| = m_i} \left[
            \sum_k {K \choose k} p_{\vec{s}}^k (1 - p_{\vec{s}})^{K - k}
        \right] \\
    &\enskip = \sum_k {K \choose k}
            \expect_{\vec{s} \text{ s.t. } |\vec{s}| = m_i} \left[
                p_{\vec{s}}^k (1 - p_{\vec{s}})^{K - k}
            \right].
 \end{split}
\end{equation}
Thus, if we wish to compute likelihood functions for $K$ shots at each sequence, we must be able to compute the $K$th moment of the distribution of sequence probabilities over all sequences of a given length.

This makes it clear how both the techniques of \citet{GranadeAcceleratedrandomizedbenchmarking2015} and \citet{HincksBayesianInferenceRandomized2018} operate.
The former restricts attention to the case in which $K = 1$, such that the needed moment is precisely that given by \citet{MagesanCharacterizingQuantumGates2012}, while the latter introduces additional parameters (formally, nuisance parameters) to track the higher moments of distributions over sequences.

Though both of these approaches are provided along with software implementations, they may be practical constraints that prevent using the $K = 1$ experimental limitation or introducing large numbers of nuisance parameters.
In practice, therefore, convenience often demands deviating from statistical principle and exploring what can be done with \emph{ad hoc} methods.
For example, least-squares methods are often used in experimental papers to report results from randomized benchmarking observations \cite{Barends2014}.
In this case, such methods are \emph{ad hoc} in the sense that least-squares fitting requires additional assumptions that are often left implicit.

In particular, if one is attempting to learn the argument $x$ of a function $f(x)$ from samples $y_i = f(x_i) + \epsilon_i$ where $\epsilon_i \sim \mathcal{N}(0, \sigma^2)$, then the least-squares solution can be readily shown to be the maximum likelihood estimator for $x$.
Thus, if a minimum variance unbiased estimator exists for $x$, it is equal to the least-squares solution.
Applying this argument to the RB case thus demands a strong additional assumption be made, namely that
\begin{align}
    n(m_i) \sim \mathcal{N}(P(m), \sigma^2).
\end{align}
By using heteroscedastic least-squares fitting, we can relax this assumption such that the variance on each $n(m_i)$ is a function of $m_i$,
\begin{align}
    n(m_i) \sim \mathcal{N}(P(m), \sigma_i^2).
\end{align}
In order to apply heteroscedastic least-squares fitting, we must therefore be able to assume normality, and we must have a way to compute $\sigma_i^2$ for each $m_i$.

In typical experiments, we do not have direct access to such variances.
That said, when synthesizing RB data from a posterior over operational representations, something remarkable happens: we can interpret the variance as the mean Bayes risk for the prediction loss over sequences.
This interpretation makes it possible to directly compute $\sigma_i^2$ from our posterior uncertainty, motivating the use of heteroscedastic least-squares fitting.

\end{appendices}

\end{document}